# Disentangling the molecular nature of the benchmark molecule $H_2O_2$ through many-body fragmentation dynamics

M. Nrisimhamurty, Anuvab Mandal, and Deepankar Misra*

Tata Institute of Fundamental Research, 1, Dr. Homi Bhabha road, Navy nagar, Colaba, Mumbai-400005, INDIA

**Abstract**

Universality in fragmentation has been the framework of scientific interest from a considerable period to disentangle the structure and dynamics of atoms and molecules. Studies in this direction however are still elusive to explore the details about polyatomic molecules because of their complexity in structure and electron-electron correlations. Hence many-body fragmentation dynamics of polyatomic hydrogen peroxide ($H_2O_2$) that plays a decisive role in atmospheric physics and chemistry and indeed which has been discovered recently in interstellar medium, is reported here. The three-body dissociation of $H_2O_2^{3+}$ which forms the main subject of this article discusses here particularly about the sequential and non-sequential fragmentation pathways and *cis*- and *trans*-isomerism that takes place in $H_2O_2$. In addition to these details, the article also includes four-body dissociation dynamics of $H_2O_2^{4+/5+}$ that has been employed to investigate the structural-chirality in $H_2O_2$.

# Main

Atoms and molecules are the building blocks of matter that make up the universe. In fact, molecules that are chiral in nature have been believed to be playing a predominant role from an astrophysical aspect and a biological relevance. Most importantly, they have been predicted to be vital in unraveling the concealed information about the origin of ubiquitous *homochirality* of terrestrial life [1-3]. From a spectroscopic view, a molecule is attributed to be chiral if and only if it exists in two mirror-spatial configurations and either of them cannot be superposed on one another by a pure translation and rotation. Moreover, for a longer time, chiroptical-spectroscopic experiments like circular dichroism, Raman optical activity and optical rotatory dispersion etc. had been employed to shine light on measuring the enantiomer-sensitive observables to distinguish the mirror-image isomers. However, recent upsurge in the development of momentum-spectrometers like cold-target recoil-Ion momentum spectrometer (COLTRIMS) [4-7] have enabled the investigations on gas-phase chirality from a molecular perspective besides providing deeper insight into many other interesting fundamental problems [8-12]. Hence to directly visualize the absolute configuration of molecular conformers in the gas phase that has been non-trivial hitherto, it is here planned to address this challenge in the smallest chiral [3] and non-planar molecule, $H_2O_2$, from an established fragmentation dynamics approach using RIMS. A recent theoretical study on the asymmetry in the chirality due to unequal number density between the enantiomeric-pure samples of $H_2X_2$ (X = O, S, Se, and Te) whenever a molecule is ionized, also increases the importance of *present studies* [13].

$H_2O_2$ is basically an asymmetric-prolate [14-16] rotor molecule that behaves as a torsional-system with a double-minima potential due to hindered internal-rotation. Torsional-mode ($\upsilon_4$) also relaxes O-H bonds into two more structural-configurations *i.e.*, *cis-* and *trans-*$H_2O_2$ [17]. It also have other vibrational-modes, namely, O-O stretching ($\upsilon_2$), O-H bending ($\upsilon_3, \upsilon_6$) and O-H stretching modes ($\upsilon_1, \upsilon_5$). Therefore it has been identified as a benchmark molecule to model atmosphere of exoplanets, cool stars and other (hot) bodies [15]. Also, it has been identified to be a prominent molecule in understanding the chemistry of ozone [18, 19], and radiation damage of biological matter [18, 20]. Besides all this, it has also been detected in the interstellar medium [21, 22] and also plays important role during (surface) chemical reactions in the formation of water, oxygen and $H_2O_4$ [23, 24]. Although the above factors highly demand kinematically-complete information about $H_2O_2$, experiments describing its structure and dynamics are scanty due to its instability in ambient conditions. Hence, present observations on dissociation dynamics of $H_2O_2$ are of immense importance. However, most importantly, the molecular nature of the $H_2O_2$ which remains as an inevitable challenge during an ion-induced polyatomic molecular-breakup, has only greatly motivated us in the current context. Hence RIMS technique, Dalitz plot and Newton diagrams [8, 9] which together has been potentially proved to the best ones in disentangling the fragmentation routes, has been employed to probe the structure and dynamics of $H_2O_2$ during its collision with 1 MeV $Ar^{8+}$. Indeed, in parallel, this article strongly emphasizes that the Newton diagrams and Dalitz plots are definitely new prospects to look at the relation between the structure and dynamics of geometrical-isomers, irrespective of any probe. To the best of our knowledge, this work will be unequivocally the first report on the signature of structural-chirality in $H_2O_2$ from the fragmentation dynamics following multiple-ionization processes, apart from the exploration of traditional fragmentation routes in a polyatomic-$H_2O_2$.

During post-experiment analysis, the following different dissociation channels have been identified from the measured fragment ions time-of-flight (TOF)-TOF coincidence mapping.

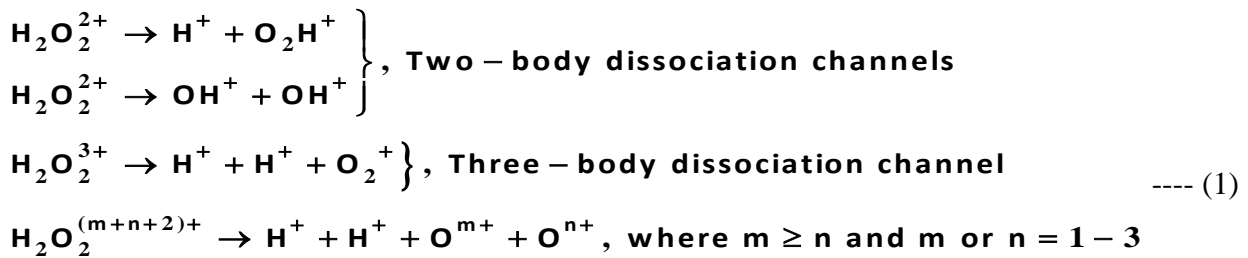

$$\left.\begin{aligned}&H_2O_2^{2+} \rightarrow H^+ + O_2H^+ \\ &H_2O_2^{2+} \rightarrow OH^+ + OH^+\end{aligned}\right\}, \text{Two}-\text{body dissociation channels}$$

$$H_2O_2^{3+} \rightarrow H^+ + H^+ + O_2^+ \Big\}, \text{Three}-\text{body dissociation channel}$$

$$H_2O_2^{(m+n+2)+} \rightarrow H^+ + H^+ + O^{m+} + O^{n+}, \text{where } m \geq n \text{ and } m \text{ or } n = 1-3$$

---- (1)

Corresponding to all these dissociation pathways, kinetic energy (KE) of the individual fragments, total kinetic-energy release (TKER) and angular distributions has been obtained from the measured momenta of coincidentally-detected fragment ions. However mainly, three- and four-body dissociation dynamics of $H_2O_2$ is alone the subject of discussion in the present context.

With the aid of coincidence measurements, it is found that in a complete-fragmentation route, $H_2O_2^{3+}$ dissociates into only one-channel with $H^+$, $H^+$ and $O_2^+$ as fragments. TKER and KE distributions shown in figure (1a) moreover corroborate that $H_2O_2^{3+}$ predominantly dissociates in a concerted-fashion, as most of the KE has been mainly shared in between the protons just as in linear molecules [**8, 9, 25-27**], by leaving behind $O_2^+$ with very low momentum.

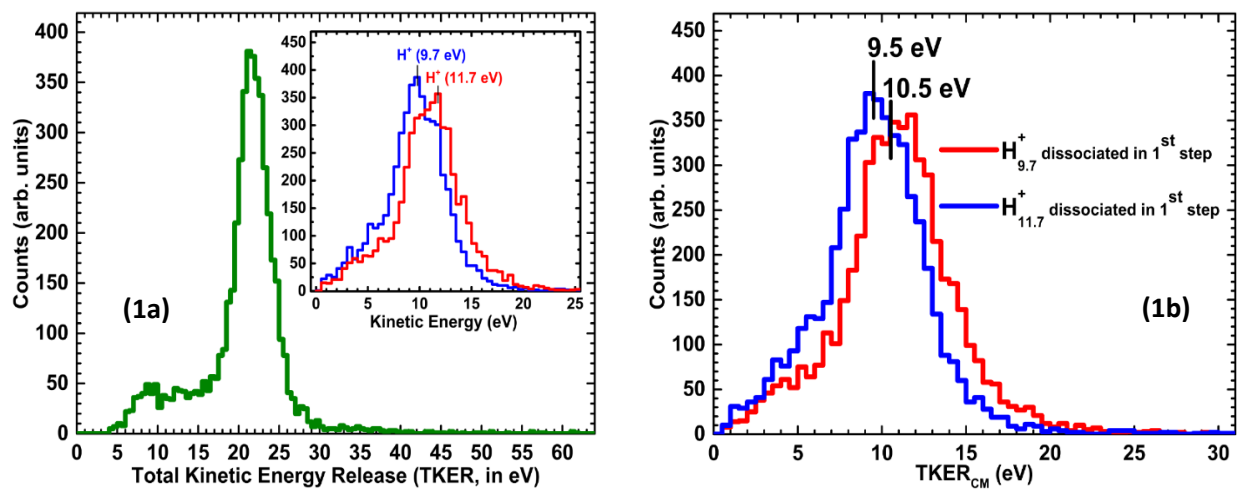

Figures (1a) and (1b) corresponds to TKER distributions in the center-of-mass (CM) frame for the concerted and sequential-decay processes. Inset of figure (1a) shows CM-frame KE distributions of either of the protons.

Now comes the point whether sequential dissociation is equally significant or not during $H_2O_2^{3+}$ breakup. If so, how to disentangle this information from the dominant direct-fragmentation mechanism, is the subject of discussion here? Generally during sequential-dissociation process, one of the O-H bonds breaks first following which second O-H fragments after a time-delay as in the equation-(1). Gradually, during the second step, $O_2H^{2+}$ which gained some angular momentum in the first breakup will rotate in addition to dissociation of $O_2H^{2+}$ into $H^+$ and $O_2^+$ fragments. This hidden two-step breakup information could also be understood from the measured momenta of the fragment ions and their respective KE deduced by considering one of the indistinguishable protons as the first fragment and the remaining $O_2H^{2+}$ as second fragment which dissociates during second-step. From the figure (1b), it is clear that derived CM-frame TKER-distributions when one among the two indistinguishable protons is dissociated during first step of sequential mechanism, ranges mostly from 2-20 eV.

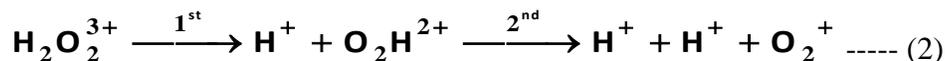

$$H_2O_2^{3+} \xrightarrow{1^{st}} H^+ + O_2H^{2+} \xrightarrow{2^{nd}} H^+ + H^+ + O_2^+ \quad \text{-----} (2)$$

To visualize kinematically-complete picture of $H_2O_2^{3+}$ breakup, Newton diagrams and Dalitz plots have been employed in this study. The particulars of these methodologies are detailed in *methods* section. For simplifying our understanding, constraints have been imposed on the

Newton diagrams and Dalitz plots w.r.t $KE_{H^+_{9.7}}$ - $KE_{H^+_{11.7}}$ correlation distribution (figure (2a)). For the ease of discussion, $KE_{H^+_{9.7}}$ - $KE_{H^+_{11.7}}$ correlation plot (figure (2a)) has been segregated into three regions like 'A', 'B' and 'C'. Here, region 'A' (*trapezoidal-block*) corresponds to a situation where $KE_{H^+_{9.7}}$ ranges from 0-8 eV and $KE_{H^+_{11.7}}$ varies from 0-9 eV. Similarly, region 'B' (*full-oval*) spans the area that spreads over 8-15 eV and 9-13 eV along $KE_{H^+_{11.7}}$ and $KE_{H^+_{9.7}}$ axes, respectively. The third and final region 'C' comprises rest of the distribution represented by *dashed-rectangular block*. Here, $KE_{H^+_{11.7}}$ covers the region under 16-40 eV and 3-10 eV in correlation with 0-13 eV and 13-26 eV along $KE_{H^+_{9.7}}$ axis, respectively. Figure (2b) shows schematic diagram depicting various points on the Dalitz plot that corresponds to different momentum-momentum correlations among $H^+$, $H^+$, and $O_2^+$. This schematic helps in understanding the possible sources for the dissociation and various breakup mechanisms that arise from the different molecular structures of $H_2O_2$. Fully fragmented three-body Newton diagram and Dalitz plot corresponding to the case with no conditions other than momentum-conservation filter is presented in figure (2c) and figure (2d), respectively. In relation with the KE-gated regions 'A', 'B' and 'C', Newton and Dalitz diagrams have been plotted accordingly in the figures (2e), (2g), (2i) and in the figures (2f), (2h) and (2j). By the definition of Newton diagram, semi-circles have been considered as a signature of two-step breakup process if and only if, the lifetime of $O_2H^{2+}$ is more than (or at least of the same order) the half-rotational time period. In Dalitz plot, the sequential decay is mainly indicated by the distribution along the straight lines parallel to each side of the triangle [**8**].

Coming to the discussion on $H_2O_2^{3+}$ breakup, a Newton diagram has been plotted in the figure (2e) with a gating corresponding to region 'A' of figure (2a). Clearly, figure (2e) show semi-circular structures along with crescent-like intense blobs, which therein infer the existence of sequential-dissociation pathway in accordance with literature [**8**]. In addition, it is also evident that the momentum distributions of the proton ($H^+_{9.7}$) and oxygen molecular-ion are seen to be off-centered on the circle defined by about half of the momentum of reference proton ($H^+_{11.7}$), which is also unambiguously a character of sequential fragmentation. However the respective Dalitz plot mapped in figure (2f) appears to be almost uniform although there is some indication of data points being relatively dense along the vector pointing towards the $O_2^+$ ion. Therefore from the Newton diagram, one can infer that there is a substantial role of two-step dissociation w.r.t region 'A' although it is not very conclusive to say from Dalitz plot.

In a similar fashion, Newton diagram and Dalitz plot are plotted in connection with the situation depicting region 'B' in figures (2g) and (2h), respectively. In comparison with region 'A', concerted process leaves behind a very strong signature for the chosen KE-KE condition and whereas the footprint of sequential breakup is negligibly small, in this case.

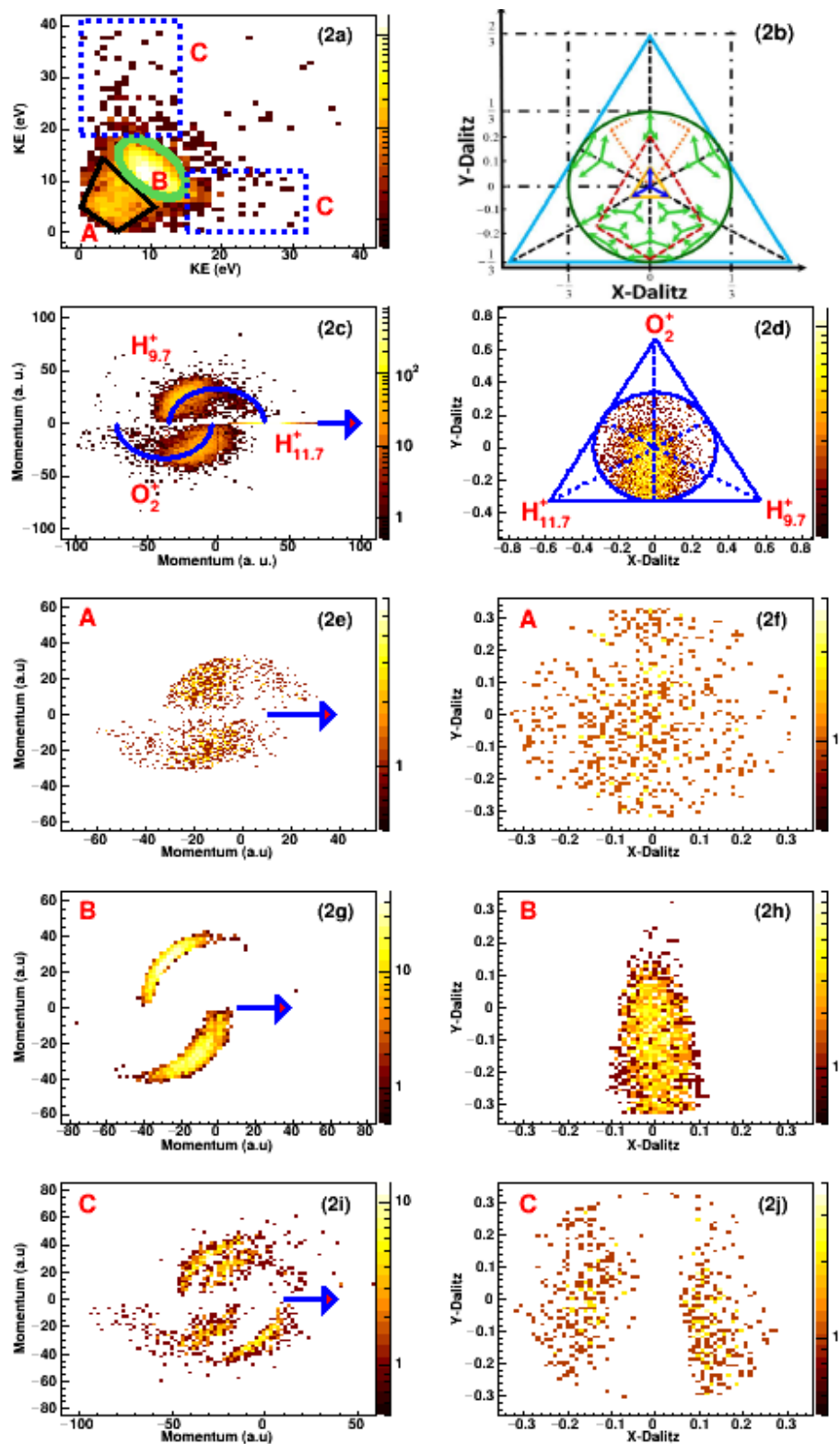

Figure (2): (a) KE-KE correlation diagram. The regions A (trapezoidal block), B (full oval) and C (dashed rectangular block) shown in the figure are selected to unveil the physics of dissociation by plotting KE-filtered Newton and Dalitz diagrams representing $H_2O_2^{3+}$ breakup. (b) Dashed lines in the schematic 3-body Dalitz plot depict the source of the dissociation from different molecular structures. Figures (c), (e), (g), (i) and figures (d), (f), (h) and (j) corresponds to the Newton and Dalitz diagrams for no condition case and w.r.t the regions A, B and C, respectively.

Newton diagram and Dalitz plot concurrent with the last and third case (region 'C') is shown in figure (2i) and figure (2j), respectively. From the Newton diagram, it is apparent that there are two semi-circles spanning over different-KE regions on either side of central $H^{+}_{11.7}$ axis. The Dalitz plot drawn in accordance evinces two linear traces crossing each other at $Y_D=0.1$ and $Y_D=0.33$. These linear traces seen in Dalitz plot as well as semi-circles of Newton diagram strongly conclude that one among the two protons is fragmented in the first-step, eventually confirming the nature of sequential dissociation during $H_2O_2^{3+}$ breakup. From these results, it is self-explanatory that sequential dissociation is equally prominent along with non-sequential breakup for the entire TKER region unlike features seen in $CO_2^{3+}$[ **8, 25, 26**] and $C_2H_2^{3+}$ [**28**] where major contribution to sequential dissociation is either only from higher or lower KER region, respectively.

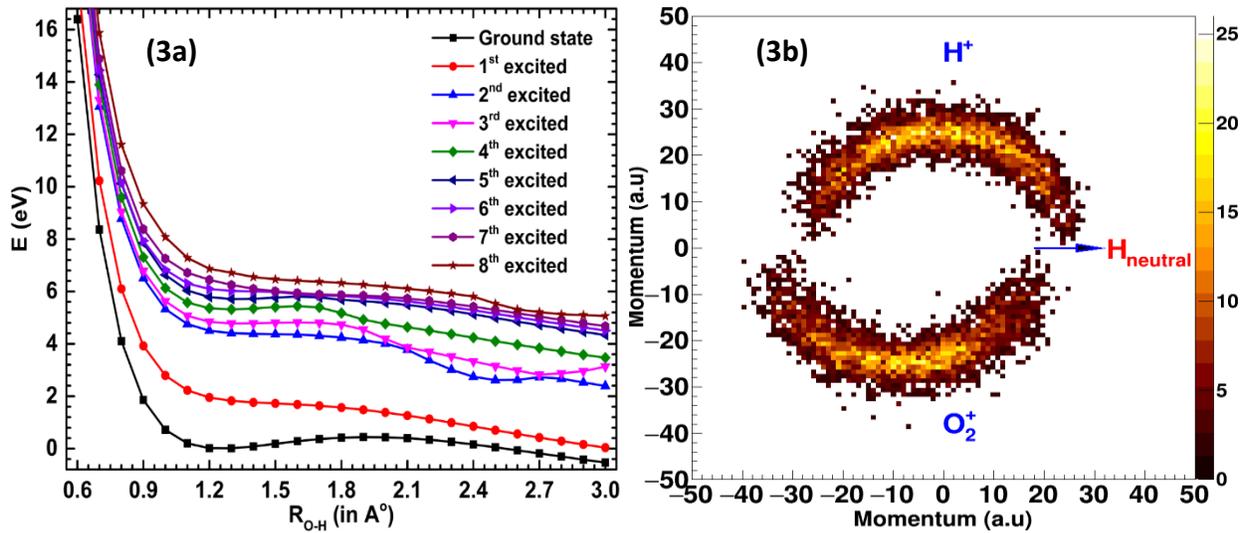

Figure (3): (a) Calculated theoretical potential energy curves (PECs) of $O_2H^{2+}$ which confirms the existence of metastable $O_2H^{2+}$. (b) Newton diagram of $H_2O_2^{2+}$ which breaks up into neutral H, $H^+$ and $O_2^+$.

In addition, quantum-chemical calculations that have been performed also conclude the existence of rotating intermediate, $O_2H^{2+}$, in metastable state. Theoretical potential energy curves (PECs) generated w.r.t. O-H bond distance has been plotted in figure (3a). Further to draw a reasonable agreement, an attempt has been made to look at two-step dissociation from an incomplete-fragmentation dynamics of $H_2O_2^{2+}$ which aids in corroborating with the survival of a metastable $O_2H^{2+}$, as in the equation-(2).

$$H_2O_2^{2+} \xrightarrow{1^{st}} H_{neutral} + O_2H^{2+} \xrightarrow{2^{nd}} H_{neutral} + H^+ + O_2^+ \quad \text{------ (3)}$$

Newton diagram drawn with *neutral hydrogen* along the preferential axis (in figure (3b)) also concretely infers that $H_2O_2^{2+}$ dissociates dominantly through two-step dissociation process. Further explanations in the context of two-step dissociation from the view of complete- and incomplete-fragmentation dynamics are provided under the *sequential fragmentation analysis of supplementary section*. Therefore from the observations drawn from the KEs, Newton and Dalitz plots; slopes and shapes of the coincidence mapping and analytical calculations provided in the supplementary section, it can be illustrated that there is definitely a sizeable-role of sequential dissociation in addition with concerted-breakup during the fragmentation of $H_2O_2^{3+}$.

Now the question here is, beyond sequential or non-sequential fragmentation dynamics, "*Is there any room left in $H_2O_2$ that has not been explored from $H_2O_2^{3+}$ breakup?*" As a matter of fact, $H_2O_2$ has been identified to be the smallest chiral molecule in the universe [**3, 29**]. It also exists as *cis*- and *trans*-$H_2O_2$ [**3, 15, 16, 29**]. Besides this, unambiguously COLTRIMS has recently been proven to be a versatile technique in the investigation on 'chirality' [**30, 31**] that remained as one of the long-standing problems in physics today. Hence, the answer will be certainly *yes*.

The point to be noted here is, "how could we attribute them as c*is*-, g*auche*- and *trans*-" from the present work? Hence to get insight into intriguing *gauche-trans-cis* isomerism in $H_2O_2$, four-body picture of $H_2O_2$ has been reduced to a three-body problem resembling $H_2O$, and to employ the three-body dissociation dynamics of $H_2O_2^{3+}$. For the ease of discussion, four-body and reduced three-body pictures of *gauche*-, *trans*- and *cis*-$H_2O_2$ are schematically depicted in figures (4a)-(4c) and in figures (4d)-(4f), respectively. To unveil the concealed information about *gauche-cis-trans* isomerism in $H_2O_2$, Dalitz plots are drawn in figures (4g)-(4i) with torsional-angle dependence, which are considered to be reflections of the character associated with the various molecular geometries of $H_2O_2$. The Dalitz plots presented in figures (4g)-(4i) are constructed for the dihedral-angles spanning over $100^O$-$160^O$, $160^O$-$180^O$ and $0^O$-$20^O$, respectively, representing the scenarios of *gauche*-, *trans*- and *cis*-$H_2O_2$.

In the present work, Dalitz plot shown in figure (4g) has clearly been dominated by the nature of *gauche*-$H_2O_2$ since it acquires a V-shaped geometry as per three-body model (see, figure (4d)). In the same way, behaviour similar to linear counter-parts [**8, 9, 25-28, 32**] has been realized from figure (4h), which thereby confirms the character of *trans*-$H_2O_2$ ($\theta_{Trans} = \theta_{H^+_{9.7} - H^+_{11.7}} = \mathbf{180^o}$), when the filter set on $\theta_{H^+ - H^+}$ ranges from $160^O$-to-$180^O$. Although a signature has been observed around $Y_D$= 0.33 (see figure (4i)) supporting the existence of *cis*-$H_2O_2$, *Arrhenius calculations* have been brought-in to understand this feature as *cis*-$H_2O_2$ is potentially unstable. Details of the analysis are outlined in *methods-Arrhenius calculations section*.

Branching-ratios estimated from these *Arrhenius* calculations are $0.158$ and $294.98 \times 10^{-5}$ whereas experimentally measured are $0.135 \pm 0.006$ and $(430.519 \pm 1.569) \times 10^{-5}$, respectively, for *trans*- and *cis*-w.r.t *gauche*-$H_2O_2$. These branching-ratios clearly infer that all three different molecular geometries seen in Dalitz-plot, are present. From the branching ratios, it can be attributed that the structural-instability of *cis*-$H_2O_2$ might be the reason for the discrepancy between theory and experimental values although experimental ratios for *trans*-$H_2O_2$ are well within 10-20% w.r.t calculated ones. However here in the present work, two scenarios are highly probable regarding the fingerprint of different geometries that has been observed. One case could be that: the target $H_2O_2$ is a random or non-racemic mixture of all the conformers depending on their stability at room temperature ($25^O$C) during sample preparation. Second one is the induced-field effect on the structure of $H_2O_2$ when it interacts with $Ar^{8+}$. A reasonable agreement between the experimental branching-ratios and that from the Arrhenius calculations which has been carried out at room temperature (25°C) concretely support the former possibility regarding random mixture of three conformers, besides induced-field effect case. More information regarding these possibilities are elaborated in great detail under supplementary section, *geometrical isomerism*.

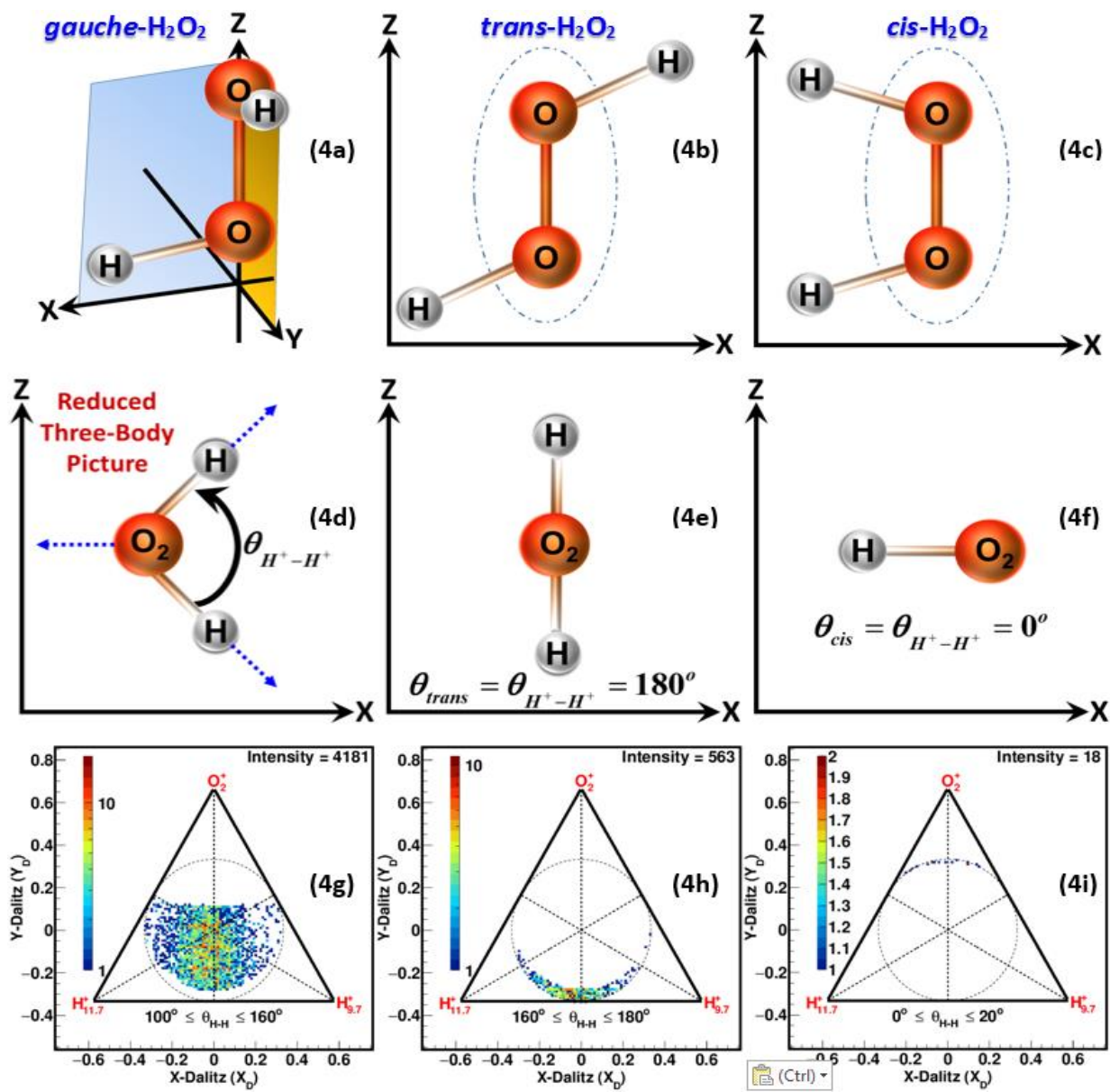

Figure (4): First row (a)-(c) corresponds to characteristic molecular-conformations, namely, *gauche-*, *trans-* and *cis-*$H_2O_2$; Reduced three-body schematic representations corresponding to the three molecular structures of $H_2O_2$ presented in first row are shown in (d)-(f) of second row. Dalitz-plots that are constructed with respect to the torsional-angle are presented in third row, from (g)-(i). *Note*: Torsional-angle ($\theta_{H-H}$) is defined by the angle between the momentum-vectors of two protons detected during $H_2O_2^{3+}$ dissociation.

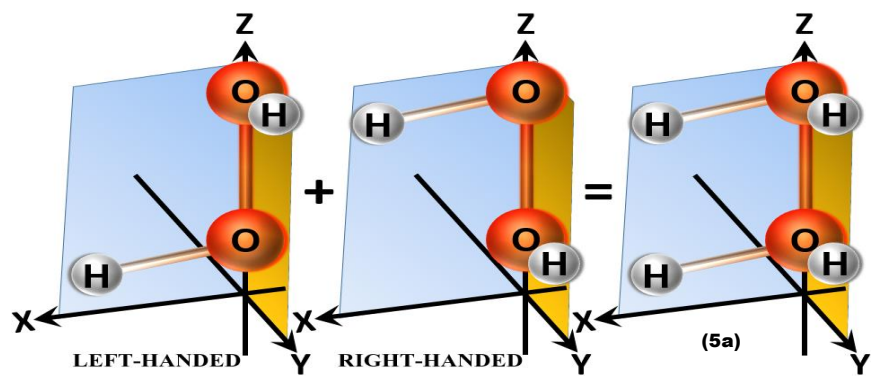

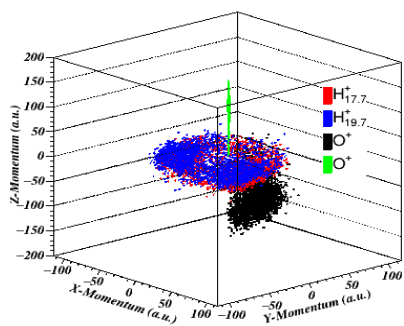
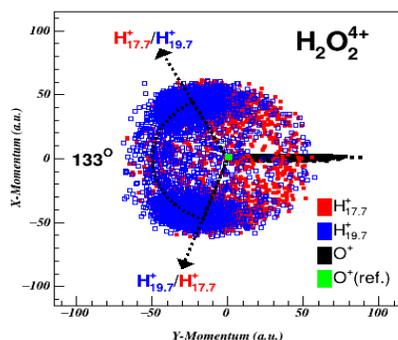

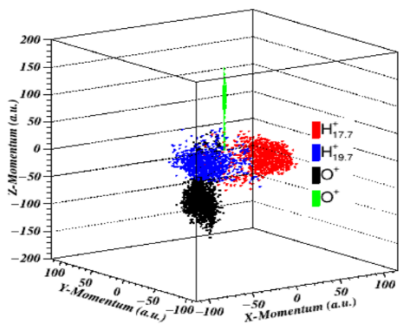
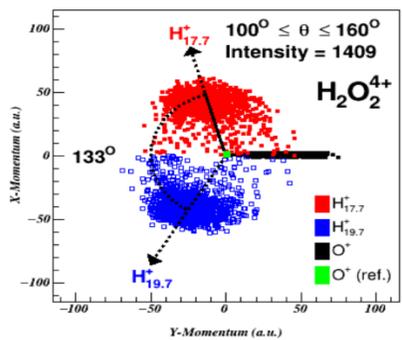

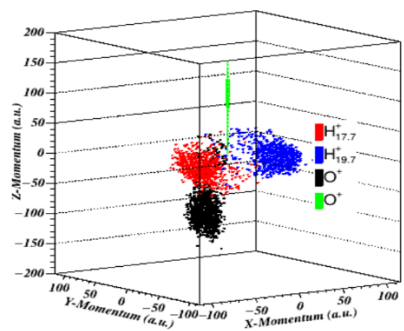
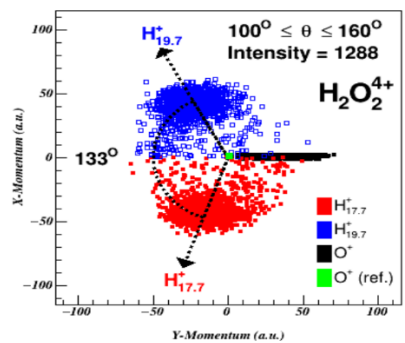

Figure (5): Schematic in (a) depicts how the experimental data is smeared out in four-body Newton diagrams that are followed by. 3d-histogram (b) corresponds to four-body Newton diagram representing $H_2O_2^{4+}$ breakup w.r.t $O^+$. Corresponding XY-projection is presented in (c). (d) depicts four-body Newton diagram of one of the chiral-isomer of $H_2O_2^{4+}$ w.r.t $O^+$ that decays into $H^+$, $H^+$, $O^{2+}$, and $O^+$ and its associated XY-projection is shown in (e). Second enantiomer 4-body Newton diagram and XY-projection are shown in (f) and (g), respectively. *Note*:Torsional-angle measured has been represented by continuous and dotted-arrows in (c), (e) and (g), for discussion here.

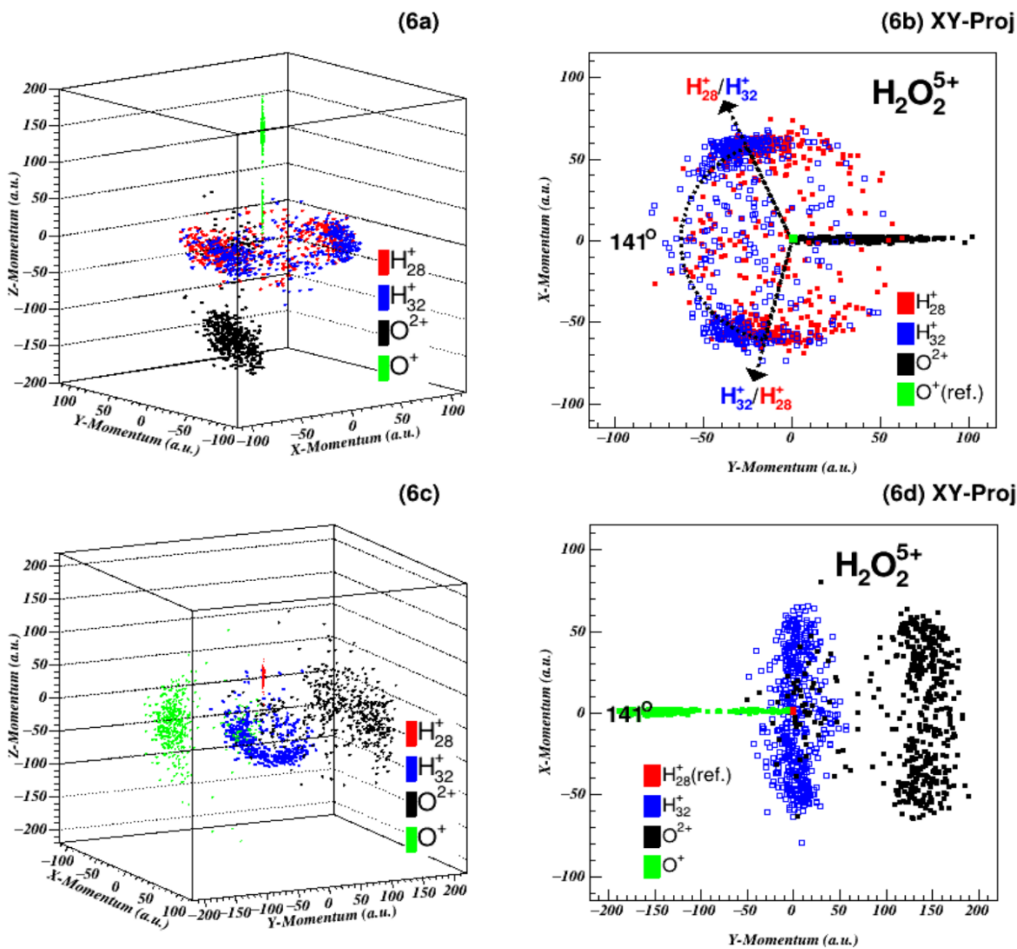

Figure (6): (a) and (b) presents four-body Newton diagram of $H_2O_2^{5+}$ w.r.t $O^+$ that decays into $H^+$, $H^+$, $O^{2+}$, and $O^+$ and its associated XY-projection, respectively. (c) and (d) shows the 4-body Newton diagram of $H_2O_2^{5+}$ and its XY-projection w.r.t $H_{28}^+$, respectively. *Note*: Torsional-angle has been represented by continuous and dotted-arrows.

Coming to the molecular-chirality of $H_2O_2$, a kinematically-complete solution has not yet been reported although there were some earlier theoretical and experimental attempts [**16, 29**]. Hence in this work, it has also been stressed on this character of $H_2O_2$ constituting left- and right-handed enantiomers, just as schematically shown in figure (5a). Four-body Newton diagrams and their corresponding XY-projections has been constructed in the figures (5b)-(6d), to comprehend this *molecular-chirality*. For the simplicity of discussion, protons denoted as $H_{17.7}^+(H_{28}^+)$ and $H_{19.7}^+(H_{32}^+)$ in the Newton diagrams has been identified with red and blue colors, respectively, whereas oxygen ions ($O^+$, $O^+$)/ ($O^{2+}$, $O^+$) with black and green (reference ion), respectively.

In a direction towards the identification of enantiomeric character of $H_2O_2$, it has been observed that the angle between the crescent-like red- and blue-blobs (figures (5c), (5e), (5g) & (6b)) is closer to $133^O$ and $141^O$, which attributes to torsional motion in the case of $H_2O_2^{4+}$ and $H_2O_2^{5+}$. For the ease of discussion, torsional-angle (see, supplementary figure (3)) has been indicated by continuous and dashed arrow-marks in XY-projections (figures (5c), (5e), (5g) & (6b)) to distinguish the nature of chirality. The results in Figure (5b) and its projection (figure (5c)) represents four-body Newton diagrams that has been plotted w.r.t $O^+$ without any constraints, is an outcome of figure (5a). The two lobes on the positive and negative X-axis in figure (5c) endorses the fingerprint of two mirror-images, although red- and blue-lobes are overlapped on one another. Hence therein to separate out the chiral-isomers, a momentum filter is set on $H^+_{17.7}(H^+_{28})$. This condition thus results in a strong signature of *right*- and *left*-oriented *gauche*-$H_2O_2$ molecules, as shown in the figures (5d) & (5e) and in the figures (5f) & (5g), respectively. The plots correspond to data presented in the figures (5d)-(5g) are conditioned on torsional angle ($100° \leq \theta_{H^+-H^+} \leq 160°$) in addition to X-momentum of $H^+_{17.7}$, to consider only the dominant gauche-$H_2O_2$. From differences in the intensities of these enantiomers, enantiomeric excess in *gauche*-$H_2O_2$ has been derived and is around 4.486%. Therefore, we strongly conclude this observed torsional-nature as a hallmark of '*molecular-chirality*' that connotes the existence of *right*- and *left*- handed *gauche*-$H_2O_2$ molecules, despite of the observed diastereomers in the earlier part of this article. A possible reason to observe this chirality, could be the asymmetry in momentum transfer (which is visible from the KE of protons), just as discussed earlier [**33**]. These results thus elucidated a clear signature of chirality in $H_2O_2$ within the framework of many-body fragmentation dynamics. To further look at the consistency of the provided information, Newton diagrams corresponding to $H_2O_2^{5+}$ have been plotted w.r.t $O^+$ and $H^+_{28}$ and is presented in the figures (6a)-(6d). Additional Newton diagrams presented in the supplementary figures (4) - (12) depending on the torsional angle $(\theta_{H^+-H^+})$ also authenticates the signature of *cis*- and *trans*-$H_2O_2$ apart from structural-chirality.

---

In this article, kinematically-complete dissociation dynamics of $H_2O_2^{n+}$ (n=3-5) has been presented. Unprecedentedly, the present results are the first experimental evidence to the best of our knowledge, in relation with enantiomers and diastereomers of $H_2O_2$ from the framework of multiple-ionization and fragmentation dynamics. Also, sequential dissociation has been observed to be on equal foot with the concerted-breakup during $H_2O_2^{3+}$ decay. The present work also identifies Newton diagrams and Dalitz plots as potential sources to discover the molecular nature (structural chirality and isomerism) in universe, irrespective of any probe. This article also corroborates with earlier reports that ascribed RIMS as an equally powerful tool to probe the microscopic configuration of simple chiral molecules. Last but not least and very important, we conclude that the present studies on $H_2O_2$, a molecule having axial-chirality, has an paramount role in the context of understanding the homochirality *i. e.*, we believe that the present work may help in accelerating the studies on the potential sources for probing the enantiomeric excess in nature. Not only that, we also state that $H_2O_2$ could be the smallest interstellar-chiral molecule in the place of recently discovered propylene oxide [CH$_3$CH$_2$CH$_2$O] [**1**]. The reason for our conclusion is: $H_2O_2$ has already been identified in the interstellar space [**21, 22**] and in addition, these results prove it can exist as chiral besides earlier results [**16**, **29**].

# Methods

Hydrogen peroxide ($H_2O_2$) molecules were prepared in a quartz tube at a room temperature ($25^O$C) and pressure typically maintained in the tube is about 0.1 Torr throughout the experiments. These molecules were allowed then to freely expand as an effusive jet into the interaction region. Well-collimated 1 MeV $Ar^{8+}$ (1 a.u, projectile vel.) highly-charged ion beam that was obtained from the Electron-Cyclotron Resonance Ion Accelerator (ECRIA), available at TIFR, Mumbai, has been used as 'projectile'. This $Ar^{8+}$ and effusive jet of $H_2O_2$ molecules were crossed at right angles at the center of the pusher and puller electrodes of the recoil-ion momentum spectrometer (RIMS). The fullest details of the RIMS were detailed elsewhere [34]. The electrons ejected after the interaction between $Ar^{8+}$ and $H_2O_2$ were extracted in the opposite direction and collected by channel-electron multiplier (CEM), will serve as the 'START' of the coincidence measurement whereas the fragmented ions do serve as the 'STOP' signal. Typical ion count rate throughout the experiment was between 700-800 Hz. All the ionic fragments produced during the interaction were guided onto a position-sensitive micro-channel plate detector by homogeneous electric fields of about 173 $Vcm^{-1}$ (ionization region) and 250 $Vcm^{-1}$ (acceleration region), respectively. A RoentDek Handels GMBH [See, 35] fast analog-to-digital converter (fADC) was used to record the data. All the data were recorded in event-by-event mode in list-mode-file(s). With the measurement of the ion time-of-flight (TOF), the detector-hit positions (x, y), geometry of the setup and electric-fields, the three-dimensional momentum vectors for each fragment ion were obtained in coincidence, in the laboratory frame. Base pressure of about $2\times10^{-8}$ Torr was maintained in the interaction chamber whereas during experiments, the operating pressure was around $2\times10^{-7}$ Torr.

For each event, momentum sum of all the ions has been calculated. This has been done to avoid background or false coincidences with a gating on the ion sum momenta. Simply for discussion, specifically two necessary and sufficient conditions has been levied on the results for complete analysis in this article.

1) The primary condition is "all the fragmented ions under study are to be detected in coincidence i.e., they should be from the same single-event" and

2) Secondly, a gating is set on the 'momentum-sum ($P_{sum}$) in each direction' depending on all fragment ions, $p_{sum-j} = \sum_{n} p_{nj}$ (n = number of fragments and j = x, y and z), to satisfy conservation of momentum. This is a reasonable approximation, if one considers that the magnitude of the momentum carried away by the ejected electrons is much smaller than that of the fragment ions.

**Newton Diagram**

**3-body**:

For the ease of discussion using 3-body Newton diagrams, the momentum vectors are assigned and presented w.r.t the center-of-mass (CM) of the constituent-fragment ions formed. All the Newton diagrams are plotted by considering the momenta corresponding to one of the two protons ($KE_{peak}\sqcup 11.7$ eV) along the x-axis and momenta of the other proton (with $KE_{peak}\sqcup 9.7$ eV) and oxygen molecular-ion fragment are chosen along the positive and negative y-axis, respectively. To simply say, the momentum vectors of $O_2^+$ and the other $H^+$ ($KE_{peak}\sqcup 9.7$ eV) has been normalized to the length of the primary proton momentum vector ($KE_{peak}\sqcup 11.7$ eV) in the upper and lower half of the plot, respectively. The length of the arrow mark indicates the most-probable value of the momentum distribution of $H_{11.7}^+$ (proton with peak KE= 11.7 eV) ions along the x-axis. The set of equations for Newton diagram corresponding to 3-body dissociation dynamics are as follows:

$$P_{1x} = \left|\vec{P_1}\right|, \quad P_{2x} = \frac{\vec{P_1} \cdot \vec{P_2}}{\left|\vec{P_1}\right|}, \quad \text{and} \quad P_{3x} = \frac{\vec{P_1} \cdot \vec{P_3}}{\left|\vec{P_1}\right|}$$

$$P_{1Y} = 0, \quad P_{2Y} = \sqrt{\left|\vec{P_2}\right|^2 - P_{2x}^2}, \quad \text{and} \quad P_{3x} = \sqrt{\left|\vec{P_3}\right|^2 - P_{3x}^2}$$

**4-body**:

To manifest four-body Newton diagrams, one of the oxygen ions have been fixed along the Z-axis. The other oxygen ion along with the fixed oxygen ion is considered to define a plane consisting of both the oxygen fragment ions. These two oxygen ions also will define unit vector along X-axis (x). The vector normal to the unit vectors along X-axis and Z-axis defines Y-axis. In the whole analogy, both the protons are defined in the space i.e., they span along all three axes. Thus, the four-body Newton picture has been depicted as a 3D-histogram w.r.t one of the oxygen ions which is placed along Z-axis whereas all the other three ions are distributed accordingly. The following are the set of equations for the four-body Newton diagrams:

$$\hat{z} = \frac{\vec{P_4}}{\left|\vec{P_4}\right|}, \quad \hat{x} = \frac{\vec{P_3} \times \vec{P_4}}{\left|\vec{P_3}\right|\left|\vec{P_4}\right|\sin(\theta_{34})}, \quad \hat{y} = \hat{z} \times \hat{x}$$

$$P_{1x} = \vec{P_1} \cdot \hat{x}, \quad P_{2x} = \vec{P_2} \cdot \hat{x}, \quad P_{3x} = \vec{P_3} \cdot \hat{x}, \quad \text{and} \quad P_{4x} = 0$$

$$P_{1Y} = \vec{P_1} \cdot \hat{y}, \quad P_{2Y} = \vec{P_2} \cdot \hat{y}, \quad P_{3Y} = \vec{P_3} \cdot \hat{y}, \quad \text{and} \quad P_{4Y} = 0$$

$$P_{1z} = \vec{P_1} \cdot \hat{z}, \quad P_{2z} = \vec{P_2} \cdot \hat{z}, \quad P_{3z} = \vec{P_3} \cdot \hat{z}, \quad \text{and} \quad P_{4z} = \left|\vec{P_4}\right|$$

## Dalitz Plot

**3-body**:

Along with Newton diagrams, Dalitz plot which was introduced by R. H. Dalitz [36] have also been used as the indispensable tool to resolve the dissociation dynamics that occurs either through sequential or concerted pathways involved. Here, in this study, to draw the two-dimensional histogram representing the 3-body Dalitz plot, the differences in the normalized kinetic energies of the two protons ($H^+$) is plotted along the x-axis while the normalized kinetic energy of $O_2^+$ is plotted in the y-axis, such that:

$$\mathbf{X-Dalitz(X_D)} = \frac{\varepsilon_1 - \varepsilon_2}{\sqrt{3}}, \text{ and } \mathbf{Y-Dalitz(Y_D)} = \varepsilon_3 - \frac{1}{3} \text{ with } \varepsilon_i = \frac{|P_i|^2}{\sum_i |P_i|^2} \text{ and } \sum_i \varepsilon_i = h$$

Where $P_i$ is the momentum of the detected fragment ion in the CM frame, altitude ($h$) =1 and $i \in H^+, H^+, O_2^+$. The main advantage of this representation is that the phase-space density is constant *i.e.*, all structure in this type of a plot comes from the dynamics of the process instead not from the trivial final state phase-space density [8]. Generally, here in this plot, conservation of momentum requires all the data events should lie within an inscribed circle of radius $\frac{1}{3}$ whereas the conservation of energy demands all that points to be within the triangle of unit height. Due to these extra feedback, this plot can also be used to substantiate the occurrence of various molecular vibration modes like symmetric ($\upsilon_1$) and asymmetric ($\upsilon_5$) O-H stretching, O-H molecular bending ($\upsilon_3$ and $\upsilon_6$) and O-O molecular stretching ($\upsilon_2$) modes of $H_2O_2$ in addition to bond-cleavage mechanisms [8]. All the conclusions corresponding to various structural geometries are drawn based on the information presented in the figure (2b) as each point ($X_D$, $Y_D$) in the Dalitz plot corresponds to a definite geometry based on the specific momentum-correlation among the three fragments $H^+$, $H^+$ and $O_2^+$. The schematic that has been drawn in the figure (2b) also explains the behaviour connected with the regions 'A', 'B' and 'C' selected in the figure (2a). The dashed lines shown in figure (2b) infers the positions where sequential dissociation traces possibly can be seen, which is even evident in figure (2h).

**In relation with Geometrical Isomerism:**

For the understanding of geometrical isomerism, three-body Dalitz-equations have been modified with an inclusion of torsional-angle dependence as below:

$$X - Dalitz(X_D) = \frac{\varepsilon_{H_{9.7}^+} - \varepsilon_{H_{11.7}^+}}{\sqrt{3}}, \text{ and}$$

$$Y - Dalitz(Y_D) = \varepsilon_{O_2^+} - \frac{1}{3} = \left[\varepsilon_{H_{9.7}^+} + \varepsilon_{H_{11.7}^+} + \left(2 \times \left(\cos\left(\theta_{H_{9.7}^+ - H_{11.7}^+}\right)\right)\right) \times \left(\sqrt{\varepsilon_{H_{9.7}^+} \times \varepsilon_{H_{11.7}^+}}\right)\right] - \frac{1}{3}$$

Here, $\theta_{H_{9.7}^+ - H_{11.7}^+}$ represents torsional- or dihedral-angle between the momentum-vectors of both the protons detected during the three-body dissociation of $H_2O_2^{3+}$. For the sake of brevity, $\theta_{H_{9.7}^+ - H_{11.7}^+}$ is redefined as $\theta_{H-H}$.

### Sequential Analysis:

To understand the sequential breakup, a simple kinematical calculation has been carried out. For analysis, the centers of the momentum distributions of $H^+$ ($p_{H_{9.7}^+}$) and $O_2^+$ ($p_{O_2^+}$), and the offset in the centers, $\delta p$, are assumed to be functions of CM momentum, $p_{O_2H^{2+}}$, and the masses of $H^+$ ($m_{H_{9.7}^+}$) and $O_2^+$ ($m_{O_2^+}$). Depending on the above considerations, the following below relations has been derived for the momentum distributions of $H^+$ ($p_{H_{9.7}^+}$) and $O_2^+$ ($p_{O_2^+}$):

$$\mathbf{p}_{H_{9.7}^+} = \left(\frac{m_{H_{9.7}^+}}{m_{H_{9.7}^+} + m_{O_2^+}}\right) \mathbf{p}_{O_2H^{2+}} \text{ --(1)} \quad \mathbf{p}_{O_2^+} = \left(\frac{m_{O_2^+}}{m_{H_{9.7}^+} + m_{O_2^+}}\right) \mathbf{p}_{O_2H^{2+}} \text{ --(2)} \quad \delta\mathbf{p} = \left|\frac{m_{H_{9.7}^+} - m_{O_2^+}}{m_{H_{9.7}^+} + m_{O_2^+}}\right| \mathbf{p}_{O_2H^{2+}} \text{ --(3)}$$

If one closely inspects the Newton diagram presented in the figure (2c), centers of the semi-circular distributions are observed to be almost at 1 a.u and -35 a.u, respectively whereas their corresponding radii is found to be around 35 a.u. From the analysis that has been carried out, deduced $\delta p$ is found to be nearly 36 a.u. and CM momentum is measured to be about 38 a.u. for the $O_2H^{2+}$.

### Arrhenius Calculations:

To accomplish the first possibility, an Arrhenius-model based tentative mathematical analysis that has been as outlined below is employed.

$$\frac{I_{less-stable}}{I_{more-stable}} \propto e^{\frac{-\Delta E}{RT}} = \left(\frac{A}{S}\right) e^{\frac{-\Delta E}{RT}},$$

Here, $\Delta E$ is the potential-energy difference between the more-stable (*gauche-*) and less-stable (*cis-* and *trans-*) isomers. 'S' is a statistical factor defined by the number and positions of potential-minima for the $C_2$ form [$C_2$ (*gauche-*), $C_{2h}$ (*trans-*), and $C_{2v}$ (*cis-*)] of $H_2O_2$. Here, contribution to the statistical-weight (S) due to the number is 3, whereas position of the minima is proportional to the width of the potential minimum hollow [37]. Let us consider the scenario between more-stable

gauche- $H_2O_2$ and relatively less-stable trans/cis-$H_2O_2$. Since $H_2O_2$ is a polyatomic gas with $C_2$ symmetry, 'RT' has been replaced with '5/2 RT'. Thus based on these inputs, the intensity branching-ratios are obtained from the calculations and present experiments as given below:

### trans-to-gauche branching ratio:

$$\frac{\Delta E}{\frac{5}{2}RT} = \frac{E_{trans}}{\frac{5}{2}RT} - \frac{E_{gauche}}{\frac{5}{2}RT} = \frac{0.04786}{0.0642} = 0.7455$$

$$\left.\frac{I_{trans}}{I_{stable-gauche}}\right|_{calc} = \left(\frac{A}{S}\right)e^{\frac{-\Delta E}{\frac{5}{2}RT}} = \left(\frac{1}{3}\right)e^{-0.7455} = 0.1582 \; ;$$

From the measurement, $\left.\frac{I_{trans}}{I_{stable-gauche}}\right|_{expt} = \left(\frac{563}{4181}\right) = 0.135 \pm 0.006$

### cis-to-gauche branching ratio:

$$\frac{\Delta E}{\frac{5}{2}RT} = \frac{E_{cis}}{\frac{5}{2}RT} - \frac{E_{gauche}}{\frac{5}{2}RT} = \frac{0.3035}{0.0642} = 4.7274$$

$$\left.\frac{I_{cis}}{I_{stable-gauche}}\right|_{calc} = \left(\frac{A}{S}\right)e^{\frac{-\Delta E}{\frac{5}{2}RT}} = \left(\frac{1}{3}\right)e^{-4.7274} = 294.98 \times 10^{-5} \; ;$$

From the measurement, $\left.\frac{I_{cis}}{I_{stable-gauche}}\right|_{expt} = \left(\frac{18}{4181}\right) = (430.519 \pm 1.569) \times 10^{-5}$

Here, energy barriers of cis- and trans- $H_2O_2$ that have been considered for the calculations are 0.3035 eV (2460 cm$^{-1}$) and 0.04768 eV (386 cm$^{-1}$) w.r.t gauche-$H_2O_2$, respectively [**16, 38-42**].

**Quantum-Chemical Calculations:**

The potential energy curves (PECs) for $O_2H^{2+}$ as a function of O–H and O–O bond length are generated using the time dependent density functional theory (TD-DFT) using the unrestricted Kohn-Sham approach. Calculations are performed using the ORCA computational chemistry package [**43, 44**]. The Becke-3–Lee–Yang–Parr (B3LYP) hybrid exchange correlation functional is used along with several basis sets, such as, Pople type 6-21G, 6-311G basis sets as well as Dunning type correlation consistent basis sets (cc-pV*n*Z and aug-cc-pV*n*Z, *n* = D, T). The PECs calculated at the TD-B3LYP/aug-cc-pVDZ level are shown in figure (3a) of main section.

**Enantiomeric Excess:**

Enantiomeric excess is a representation to describe the relationship between the two enantiomers in a mixture. In the present work, sample chosen is a random or non-racemic mixture. Therefore intensities of either of enantiomers (chiral isomers) are quantitatively measured with the aid of torsional angle (angle between the protons). Thus, the measured enantiomeric excess has been deduced with the aid of the following below formulation:

$$\text{Enantiomeric excess (ee) (in \%)} = \left( \frac{I_{enantiomer-1} - I_{enantiomer-2}}{I_{enantiomer-1} + I_{enantiomer-2}} \right) \times 100$$

$$\text{Enantiomeric excess (ee) (in \%)} = \left( \frac{1409 - 1288}{1409 + 1288} \right) \times 100 = 4.486\%$$, where 1409 and 1288 are the intensities of the $H_2O_2$ enantiomers presented in the supplementary figures. (11) and (12).

# Supplementary Information

**Sequential fragmentation:**

To conclude further the experimental observations regarding sequential decay, a simple kinematical calculation has been performed for the $H_2O_2^{3+}$ complete dissociation channel. The particulars of the procedure are provided in *sequential analysis* of *methods* section. From the centers and radii of the semi-circles, momentum distributions of $H^+$ and $O_2^+$ are derived. These momenta thereby aided to deduce mean TKER that corresponds to the case when $O_2H^{2+}$ ion fragments into $H^+$ and $O_2^+$ during the second-step of sequential decay. This estimated TKER is found to be roughly about 9.36 a.u. and is relatively in good agreement with the mean TKER of the figure (1b) (main discussion) where $H^+$ ($H_{11.7}^+$) dissociates in the first-step of the sequential dissociation process. These calculations also acts as a substantial support regarding the sequential dissociation during fragmentation of $H_2O_2^{3+}$.

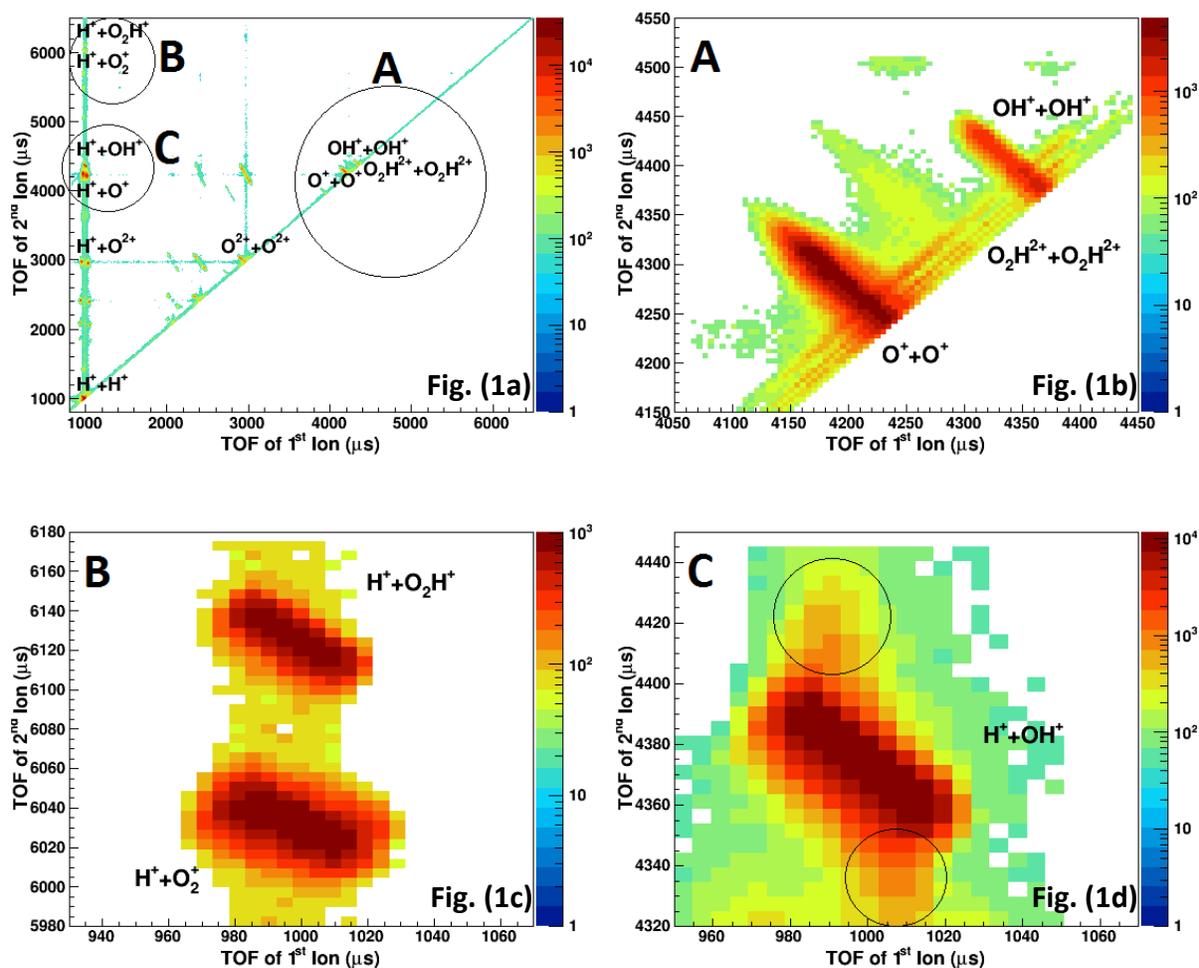

Suppl. Fig. (1): $TOF_1$-$TOF_2$ coincidence mapping diagram of 3-body dissociation of $H_2O_2$ in collision with 1 MeV $Ar^{8+}$

To visualize more details about the rotating intermediate $O_2H^{2+}$ and further understand the observations regarding two-step dissociation, $H^+$ TOF ($TOF_1$)-$O_2^+$ TOF ($TOF_2$) correlation diagram (Suppl. Fig. (1)) is plotted as dissociation mechanism can also be connoted based on the shape and slope of the ions time-of-flight (TOF)-TOF coincidence mapping. The complete ion

TOF$_1$-TOF$_2$ (H$^+$ TOF (TOF$_1$)-O$_2^+$ TOF (TOF$_2$)) correlation diagram of H$_2$O$_2^{3+}$ has been shown in the suppl. fig. (1a), whereas magnified view corresponding to encircled regions A, B and C of suppl. fig. (1a) are presented in the plots 'A' (suppl. fig. (1b)), 'B' (suppl. fig. (1c)) and 'C' (suppl. fig. (1d)).

In order to understand the information regarding the sequential decay from the shapes and slopes of TOF-TOF coincidence map, selection rules set by J. H. D. Eland [**45, 46**] have been employed. As per their rule [**45**], *sequential decay* will only be seen in the case when a neutral with low momentum fragments first leaving behind the remaining moiety with larger momentum during the first-step of incomplete-dissociation process. Hence, Hence, momentum-sum of the oxygen ion and the proton have been chosen to a value such that $abs(p_{sum-j}) \leq 10 \ a.u$ to invoke the discussion here.

From correlation diagram 'A' (suppl. fig. (1b)), it is clear that there is a channel contributing to O$_2$H$^{2+}$+O$_2$H$^{2+}$ dissociation pathway. The slopes $s_d$ and $s_i$ are deduced from the data plotted in the diagram 'B' (suppl. fig. (1c)). Diagram 'B' shows that the shape of this H$^+$+O$_2^+$ island is like a "lozenge", thereby implying that the undetected light neutral H atom has low momentum, and eventually low KE, which is what observed in the present scenario (suppl. fig. (2)) for the case of $abs(p_{sum-j}) \leq 10 \ a.u$. The shape also introduces us to an information that detected fragment ion-pair are anti-correlated on an average with random momentum components, of the same sign and of magnitude proportional to mass of each fragment ion [**45**]. From the earlier studies [**45, 46**], the slope of the island in case of initial-charge separation process ($s_i$) is given by slope, $s_i = (-1)\left(\dfrac{q_1}{q_2}\right)\left\{\dfrac{m_2}{(m_2+m_3)}\right\}$ whereas deferred-charge separation ($s_d$) is $s_d = (-1)\left(\dfrac{q_1}{q_2}\right)$.

From the present analysis of H$_2$O$_2^{2+}$ incomplete fragmentation channel, it is found that both the slopes $s_d$ and $s_i$ are close to $\approx$ **-1**, which is an essential condition for sequential dissociation as described in the previous work [**45, 46**]. In addition, TOF-TOF correlation diagram 'C' (suppl. fig. (1d)) indicates a dissociation channel H$^+$+O$_2$H$^{2+}$ by the encircled regions with definitely a different slope from that of H$^+$+OH$^+$. This infers as well as supports the information derived from the theoretical potential energy curves (PECs) regarding existence of metastable rotating intermediate O$_2$H$^{2+}$. It is also trivial to draw a conclusion that the momentum of neutral H is relatively lower than the proton and O$_2^+$ momentum from the KEs presented in the supplementary figure (2), for the momentum-difference condition chosen.

Therefore, we infer that the KER measured, Dalitz plots, Newton diagrams and slopes deduced in the current work, figures (1)-(3) of the main article and suppl. fig. (1) & (2), strongly herald a possibility for sequential dissociating channel.

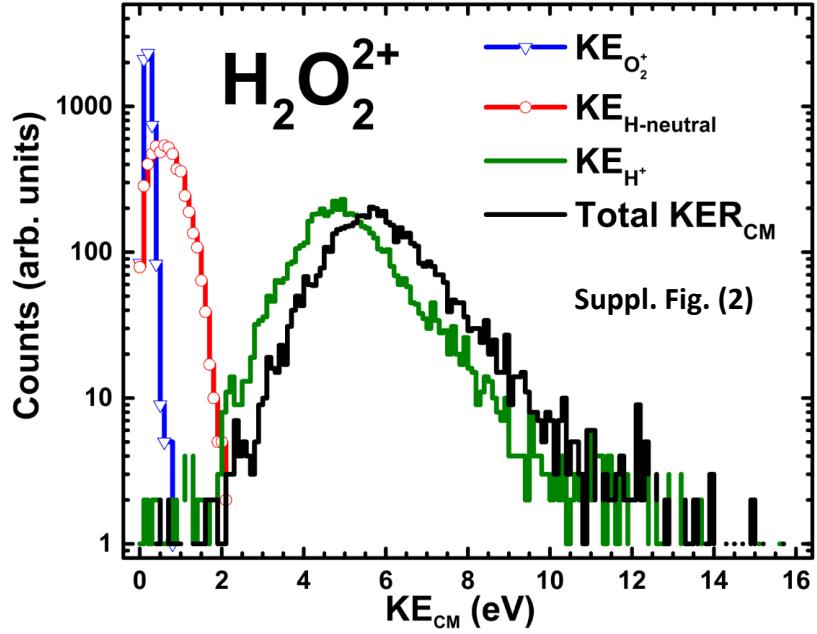

Suppl. Fig. (2): This plot presents the KEs of the neutral, proton and oxygen molecular ion that are produced during an incomplete three-body dissociation dynamics of $H_2O_2^{2+}$, for a condition on 'momentum-sum' such that $abs\left(p_{sum-j}\right) \leq 10 \ a.u$.

*Note*: In fact, neutral H produced in a two-step fragmentation shares momentum with the $H^+$ and $O_2^+$, if the neutral forms in the second-step of dissociation process. What if the neutral fragment forms in the first-step of the sequential dissociation? In that case, neutrals should ideally gain negligibly small momentum in comparison with charged ones as they are not formed from coulomb explosion. Hence, this is the reason for us to choose $abs\left(p_{sum-j}\right) \leq 10 \ a.u$ for the discussion on incomplete fragmentation'.

**Geometrical Isomerism:**

The viable reasons regarding the geometrical isomerism that has been presented in the main article is needed to be verified. For that purpose, the discussion on *geometrical isomerism* has been elaborated below from the perspective of hopping between the three conformers. Thus in a direction to understand it, the following reasons are expected to be the most likely ones to trace out the footprint of the three molecular conformers of $H_2O_2$ that has been seen in the Dalitz plots (figures ((4g)-(4i)).

(i) *Probability-1*: These three geometries that has been identified in this work could be possible, if and only if, the as-prepared gas-mixture is a random-mixture of $H_2O_2$ molecules with different conformations.

(ii) **Probability-2**: Secondly, it could also be plausible by effects-induced by the electric-field of 1 MeV $Ar^{8+}$ projectile during its interaction with $H_2O_2$. However, only *gauche*-$H_2O_2$ and *cis*-$H_2O_2$ are polar by nature [47].

To realize the above promising possibilities in the present scenario, Arrhenius-model based tentative mathematical analysis has been carried out as outlined in *methods-Arrhenius calculations section*. The tentative branching-ratios of *cis*- and *trans*- as a function of skewed-$H_2O_2$ are estimated to be $0.158$ and $294.98 \times 10^{-5}$ whereas measured branching-ratios are found to be $0.142$ and $286.26 \times 10^{-5}$, respectively. From these calculations and the details presented in the figure (4) of main section, it is clear that the inferences drawn related to the existence of three different molecular geometries based on Dalitz-plot, is true. Even though the calculated and experimental branching-ratios are little different, they are well within the 3-10% error-bar w.r.t calculated values. In fact, typical radiative lifetimes of the ro-vibrational states that spans the energy regime (cm$^{-1}$) of *cis*-, *gauche*- and *trans*-$H_2O_2$ is found to be of the order of 0.01-10 s, as per the Exomol database [15]. These lifetimes indeed will substantiate the argument regarding the formation of all the three conformers at the time of preparation of the sample. This could be attributed as such because the timescales are sufficient enough for these three geometries to sustain until when the interaction happens between $H_2O_2$ and the projectile, 1MeV $Ar^{8+}$.

Now comes the point whether these branching-ratios pave a way to conclude that the target $H_2O_2$ is a random or non-racemic mixture of *gauche*-, *trans*-, and *cis*-$H_2O_2$ according to their stability at room temperature (25$^O$C) during sample preparation, in correlation with the probability of hopping or tunneling between the three isomers at the room temperature (25$^O$C). In the present experiment, the target sample is considered in 50:50 ratio of water and liquid-$H_2O_2$. Therefore the as-prepared target is a stoichiometric sample of $H_2O$, $O_2$ and $H_2O_2$ in gas-phase because liquid-$H_2O_2$ decomposes into water and oxygen at ambient conditions. The typical vapor pressure ($\Delta P(H_2O_2)$) of $H_2O_2$ that has been maintained throughout experimental period during when fragmentation data has been recorded, is nearly about $\leq 1$ Torr. Thus, looking at the results and the sample preparation, it can stated that $H_2O$ might be acting as a bi-functional catalyst that transfers a proton/ hydrogen preceding hydrogen/proton-acceptor behaviour in the initial stage of the process, as [48] attributed earlier. It was also reported [48] that 1, 2-hydrogen migration of $H_2O_2$ transits through a very high potential barrier due to the following two factors: 1) the reaction took place could be endothermic in nature and 2) the migrating proton polarizes the O-O bond eventually driving it to break at the transition state. In the similar way, when the water (or any other polar solvent) comes into contact with $H_2O_2$ molecules during target preparation of the present experiments, the energy barrier might have been noticeably lowered (depending on the solvent) for the case of *cis*- although it remains high in comparison with *gauche*- or *trans*- whereas the *trans*- barrier might have been increased as discussed earlier [see, **47**]. Therefore as theoretically demonstrated [47], the *gauche*- and *cis*- which do have a dipole moment might become stable in the polar environment but not *trans*-$H_2O_2$ which has zero-dipole moment. Shortly to say, if the dielectric constant of the polar medium increases, the energy barrier of *cis-* conformer decreases whereas *trans-* increases. From the results of Du et al. [47] on $H_2O$-$H_2O_2$ molecular complex, *trans*-$H_2O_2$ structure attains a non-zero dipole-moment instead of being zero as in the gas-phase. However, this possibility of polar *trans*-$H_2O_2$ is very unlikely to occur as the present experiments are carried out in coincidence fashion. By taking advantage of the current results in accordance with [47, 48], it has been inferred that if non-polar *trans*-$H_2O_2$ or polar *cis*-$H_2O_2$ are formed during sample-preparation stage, they are more likely to stay stable in vacuum (or non-polar solvents), in accordance with their stability lifetimes. Therefore, it can be concluded that the target-$H_2O_2$ is a mixture of all three conformers resulting from the hopping or tunneling through the barriers between their configurations due to these solvate effects. Further one can also exemplify that the solution- and gas-phases do play an important role in stabilizing a particular isomer from tunneling or hopping through the barrier to other configuration. Therefore, it can be illustrated that the source of "*cis*-to-*gauche*-to-*trans*-isomerization" seen in the current Dalitz plots plotted from the dissociation dynamics of $H_2O_2^{3+}$ breakup can be well understood from the present experimental intensity branching-ratios and the plausible in-built solvate effects during target-preparation. This isomerism can also be identified as "Solvate-" or "Hydrate-isomerism" due to water being a solvent. These all might be the reasons for the present results to differ a lot from the case of linear-acetylene where too isomerization takes place between acetylene and vinylidene as observed earlier [10, 49] and solvate effects in the paper by Bosch *et al*. [47].

      Coming to the second probability, the structural transition might also take place due to the induce-field impact on $H_2O_2$ molecules by an energetic projectile like 1 MeV $Ar^{8+}$, as in the present case. Even though this case as well is equally probable in addition to the plausible target-preparation, *trans*-$H_2O_2$ which is a non-polar molecule in gas-phase (without solvent interaction) cannot be formed during interaction with projectile. Therefore, it can be concluded that the signature of *trans*-$H_2O_2$ seen could be most likely from the initial target prepared and not from the induced-field effects. It is also true *cis*- $H_2O_2$ becomes unstable in vacuum (non-polar) unlike the situation of polar-environment (water). This clearly evinces that although there is a possibility for *cis*-$H_2O_2$ to occur when *gauche*-$H_2O_2$ or other structures are reconstructed due to induced-field effects from projectile interaction, its contribution is infinitesimally small as it is unstable in non-polar environment. Thus, it can be attributed that this hypothesis might be a highly-efficient one for *cis*-$H_2O_2$ even though it is unstable but definitely not for *trans*-$H_2O_2$ formation although it is stable in vacuum [47]. Hence, it can be concluded that the first reason is a most probable one than induced-field effects and is also evident from the intensity-branching ratios that are deduced from experiments and calculations above and radiative lifetimes, regarding the observed *cis-/trans-*isomerism.

**Torsional-angle measurement from $H_2O^{3+}$, $H_2O_2^{3+}$, $H_2O_2^{4+}$ and $H_2O_2^{5+}$ breakup channels:**

The correlation angle between the protons that are detected in coincidence during $H_2O^{3+}$, $H_2O_2^{3+}$, $H_2O_2^{4+}$ and $H_2O_2^{5+}$ breakup dynamics has been compared here. As mentioned in the main article, $H_2O_2$ attains V-shaped geometry just like $H_2O$ when it is reduced to a three-body picture with $H^+$, $H^+$, and $O_2^+$ in the place of $H^+$, $H^+$, and $O^+$ as the ejected fragments. This is apparent from the correlation-angles plotted between the momentum vectors corresponding to either of the protons as shown below in the figure (3a). In the context of molecular-chirality as discussed in this work, the angle derived from the momentum vectors of the indistinguishable protons that are detected during $H_2O_2^{4+}$ and $H_2O_2^{5+}$ dissociation dynamics have been plotted, as in the below figure (3b). The correlation-angles have been found to be $120^O$, $133^O$, $133^O$ and $141^O$ for the cases of $H_2O^{3+}$, $H_2O_2^{3+}$, $H_2O_2^{4+}$ and $H_2O_2^{5+}$ breakup. The angles measured in the purview of three-body $H_2O_2^{3+}$, and four-body $H_2O_2^{4+}$ turns out to be same and in fact, it is to be, which further in return confirms the results presented in this work regarding enantiomer and diastereomers character.

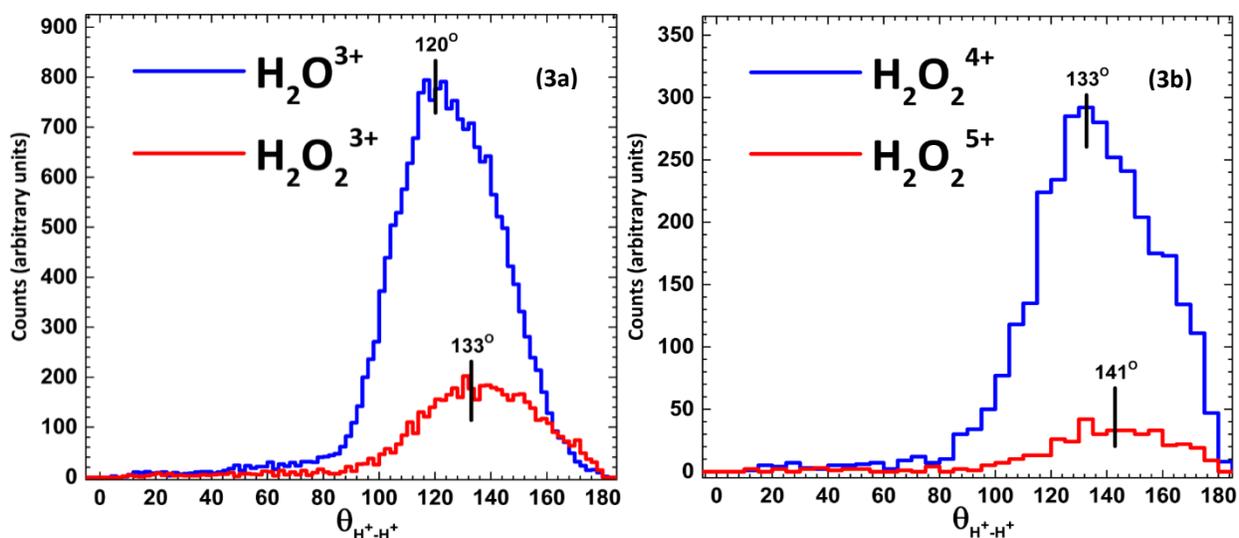

Suppl. Fig. (3): (a)-Correlation angle $H^+:H^+$ between the momentum vectors of the two protons produced from $H_2O_2^{3+}$ and $H_2O^{3+}$ breakup dynamics during 1 MeV $Ar^{8+}$ ion-induced impact, (b)-Correlation angle $H^+:H^+$ between the momentum vectors of the two protons produced from $H_2O_2^{4+}$ and $H_2O_2^{5+}$ breakup dynamics

To get qualitative and quantitative understanding about the *gauche-cis-trans*-isomerism, 3D-histograms representing the four-body Newton diagrams w.r.t torsional-angle dependence are shown in suppl. figures (3)-(10). However due to a large statistical error regarding *cis*-isomer, to visualize the real existence of *cis*-$H_2O_2$ in the present data, Newton diagrams corresponding to the cases where $0^o \leq \theta_{H^+-H^+} \leq 40^o$ and $0^o \leq \theta_{H^+-H^+} \leq 60^o$ are presented in the figures (9) and (10). These cases qualitatively resemble a structure closer to *cis*-$H_2O_2$ isomer than *gauche*-$H_2O_2$ as the torsional angle for the *gauche*-$H_2O_2$ is around $133^o$. Therefore, it can be confirmed that *cis*-$H_2O_2$ molecular geometry that has been identified in the present work, is true.

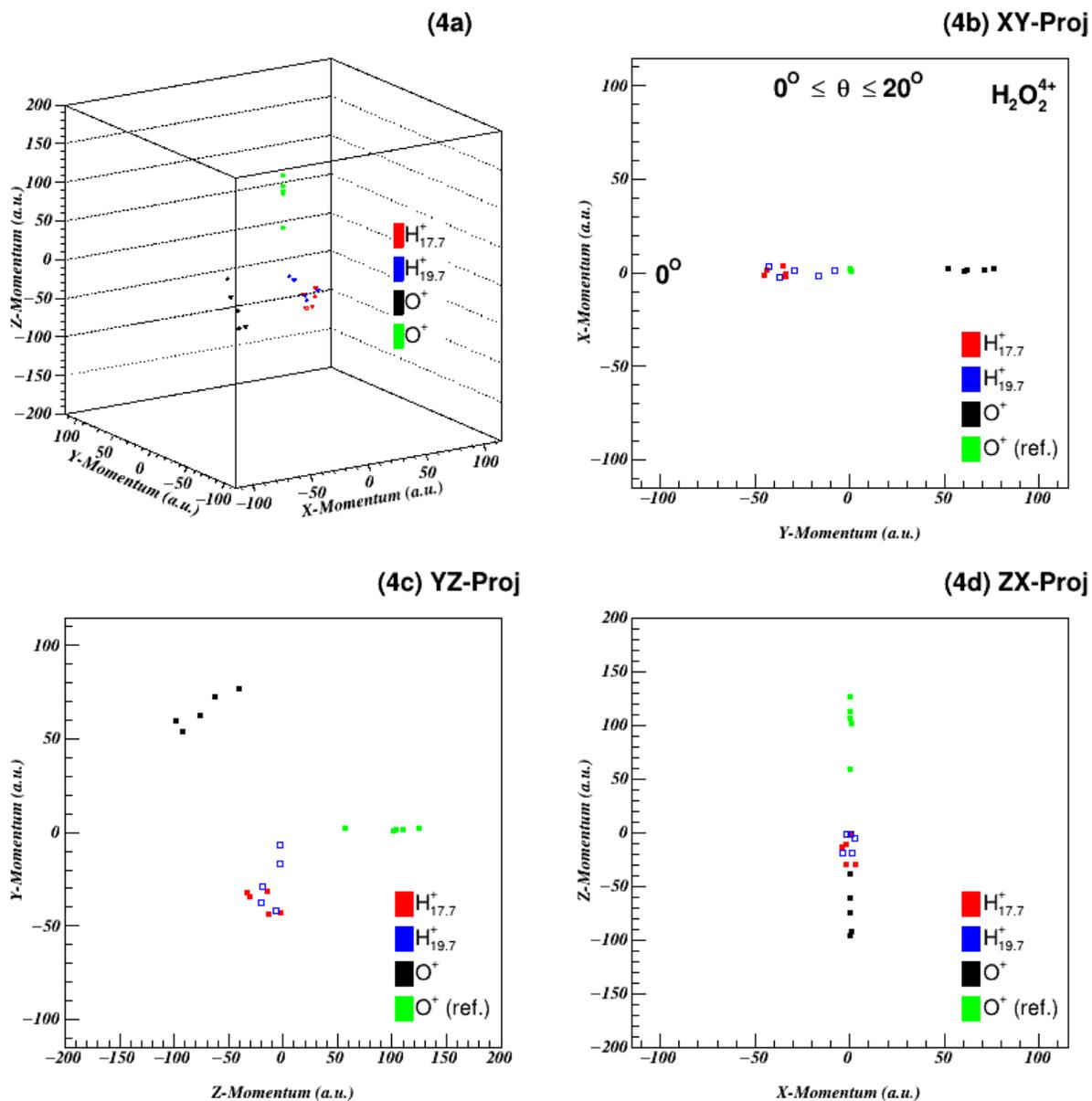

Suppl. Fig. (4): (a) Four-body Newton diagram representing the full 4-body breakup of $H_2O_2^{4+}$ for the torsional angle condition $0^o \leq \theta_{H^+-H^+} \leq 20^o$, which resembles the geometry of *cis*-$H_2O_2$. The XY-, YZ-, and ZX- projections are provided in (b), (c) and (d), respectively.

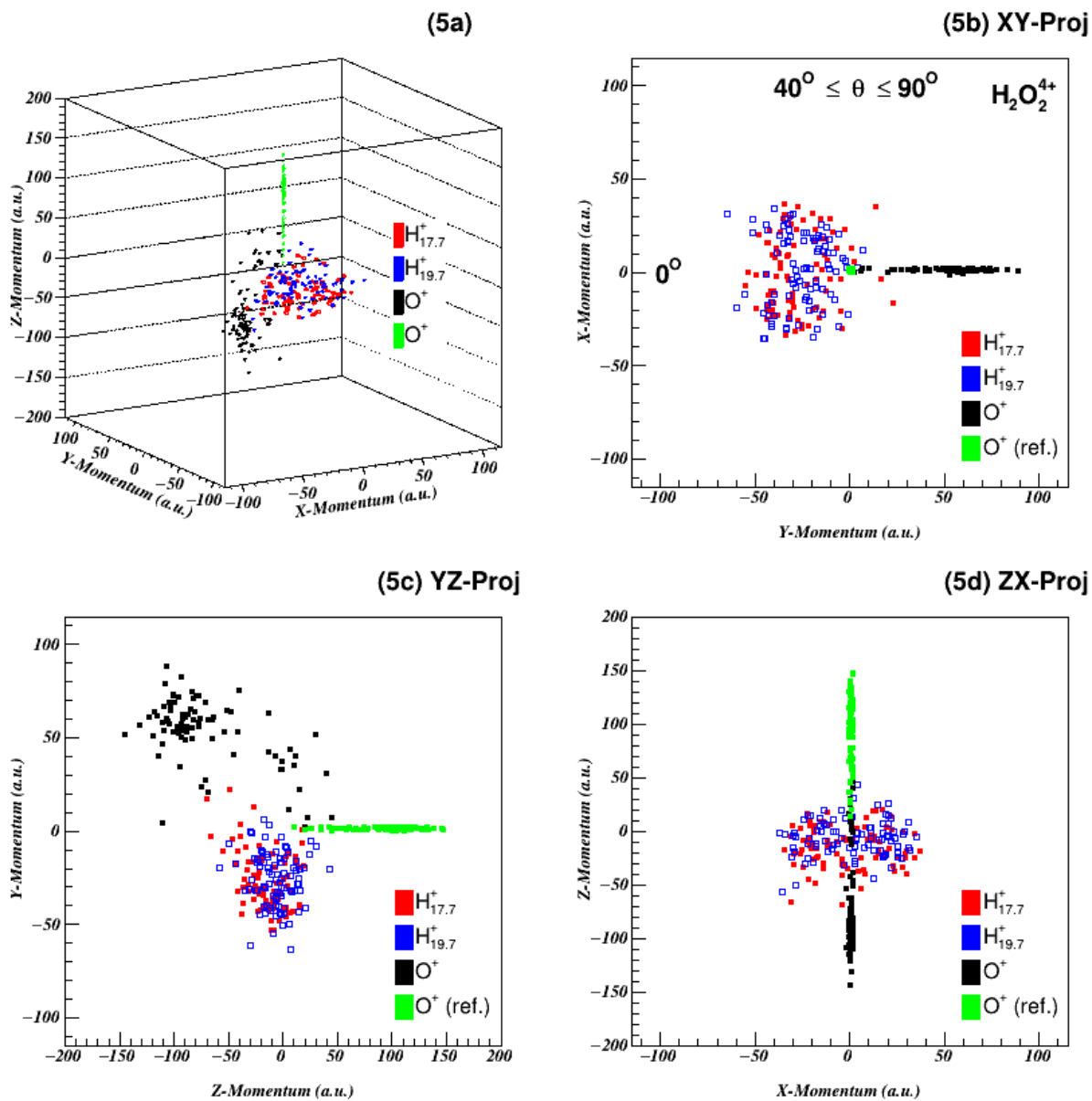

Suppl. Fig. (5): (a) Four-body Newton diagram representing the full 4-body breakup of $H_2O_2^{4+}$ for the torsional angle condition $40^o \leq \theta_{H^+ - H^+} \leq 90^o$. The XY-, YZ-, and ZX-projections are provided in (b), (c) and (d), respectively.

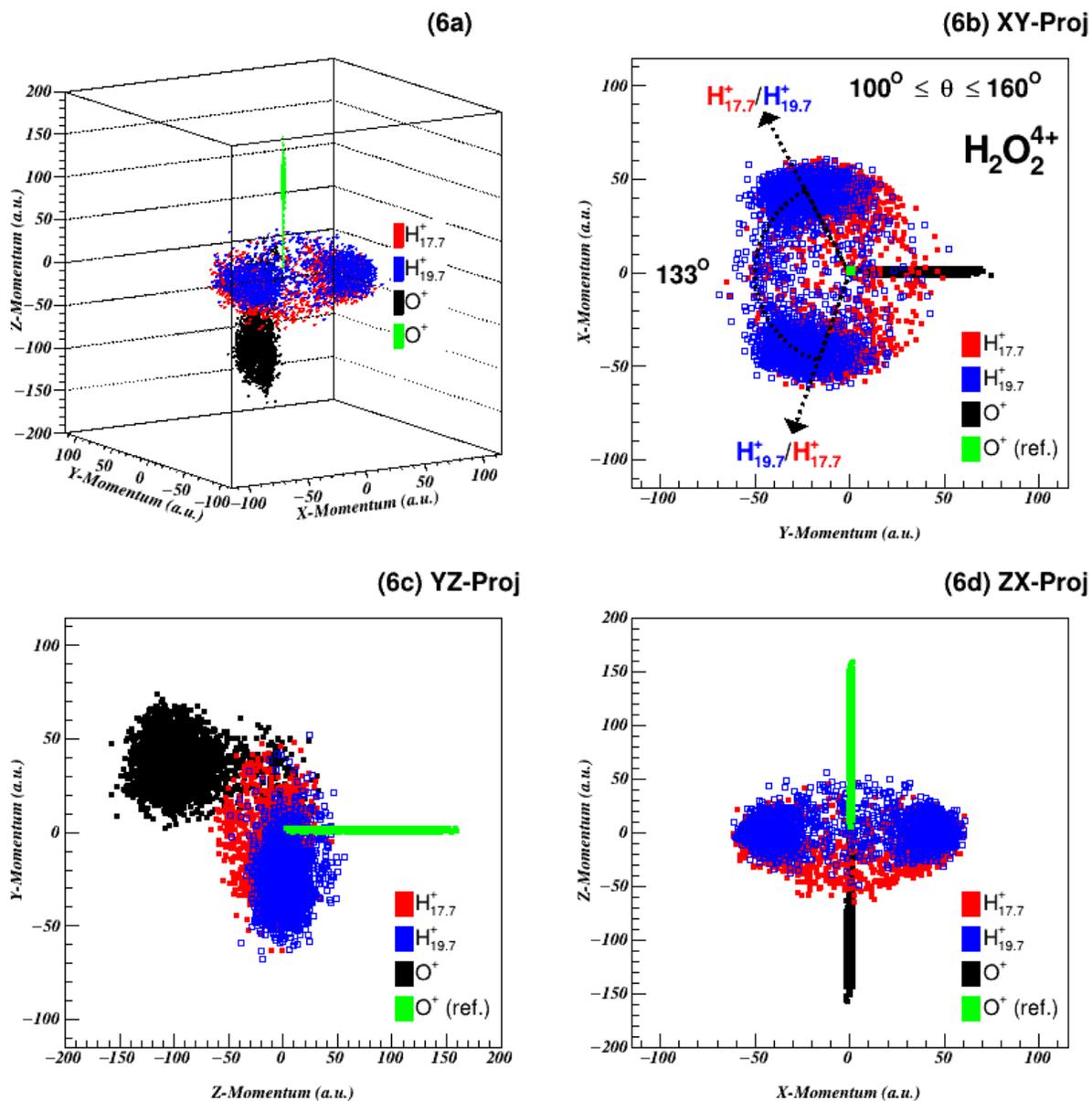

Suppl. Fig. (6): (a) Four-body Newton diagram representing the full 4-body breakup of $H_2O_2^{4+}$ for the torsional angle condition $100^o \leq \theta_{H^+-H^+} \leq 160^o$, which corresponds to *stable gauche*-$H_2O_2$ geometry. The XY-, YZ-, and ZX-projections are provided in (b), (c) and (d), respectively. The dashed-arrows indicate the structural configurations which are mirror-images of one another and non-superimposable. These arrow-marks thus basically indicates the signature of chiral isomers.

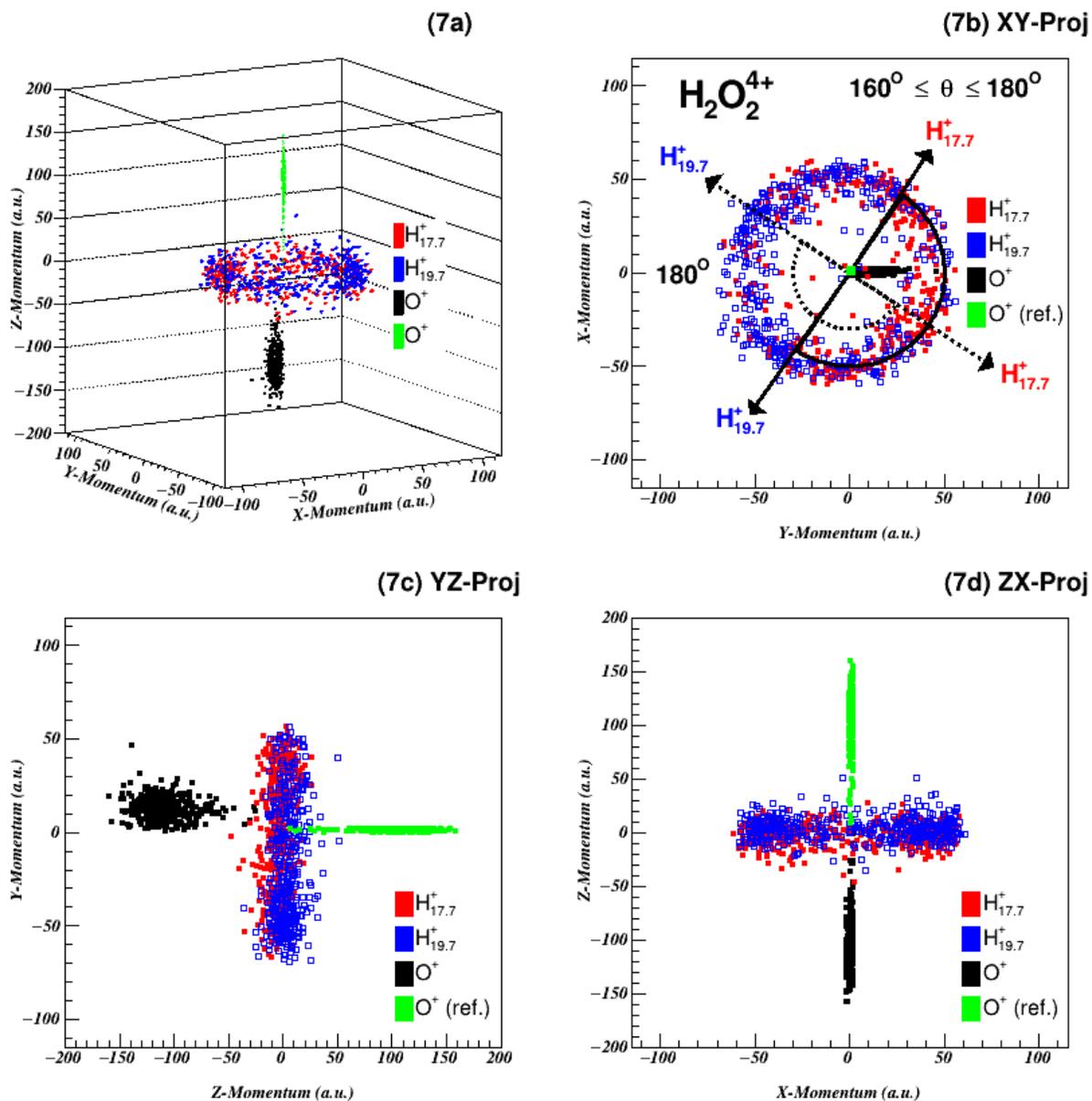

Suppl. Fig. (7): (a) Four-body Newton diagram representing the full 4-body breakup of $H_2O_2^{4+}$ for the torsional angle condition $160^o \leq \theta_{H^+-H^+} \leq 180^o$, which depicts *trans*-$H_2O_2$ geometry. The XY-, YZ-, and ZX-projections are provided in (b), (c) and (d), respectively.

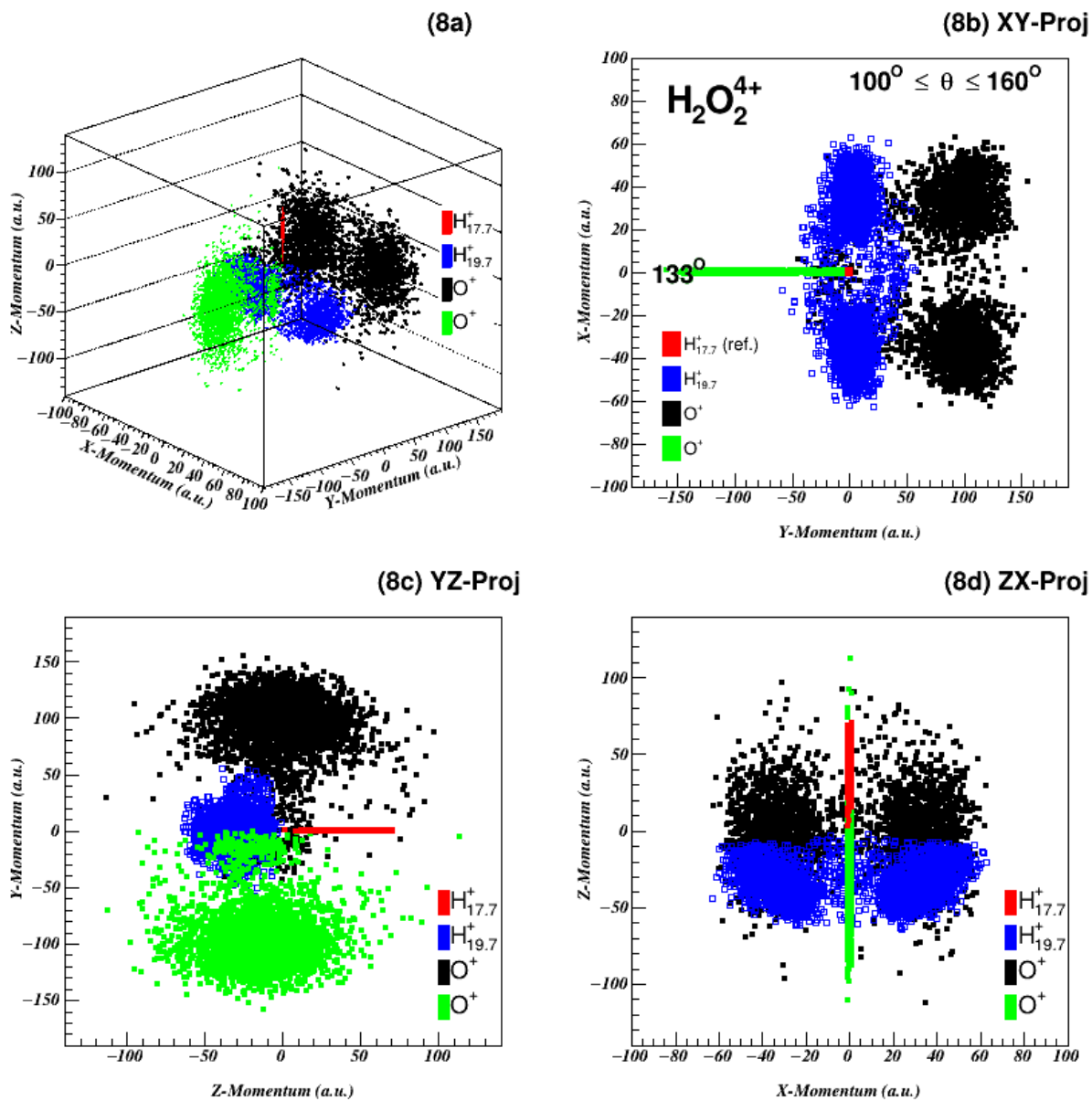

Suppl. Fig. (8): (a) Four-body Newton diagram representing the full 4-body breakup of $H_2O_2^{4+}$ for the torsional angle condition $100^o \leq \theta_{H^+-H^+} \leq 160^o$, which corresponds to *stable gauche*-$H_2O_2$ geometry. The XY-, YZ-, and ZX-projections are provided in (b), (c) and (d), respectively. The two blue lobes in the (b) indicate the chiral isomers that are mirror-images of one another and non-superimposable. Note: Here in this plot, $H_{17.7}^+$ has been considered as the reference-ion instead of $O^+$ ion (previous case) to check the consistency of the discussion on *structural-chirality*.

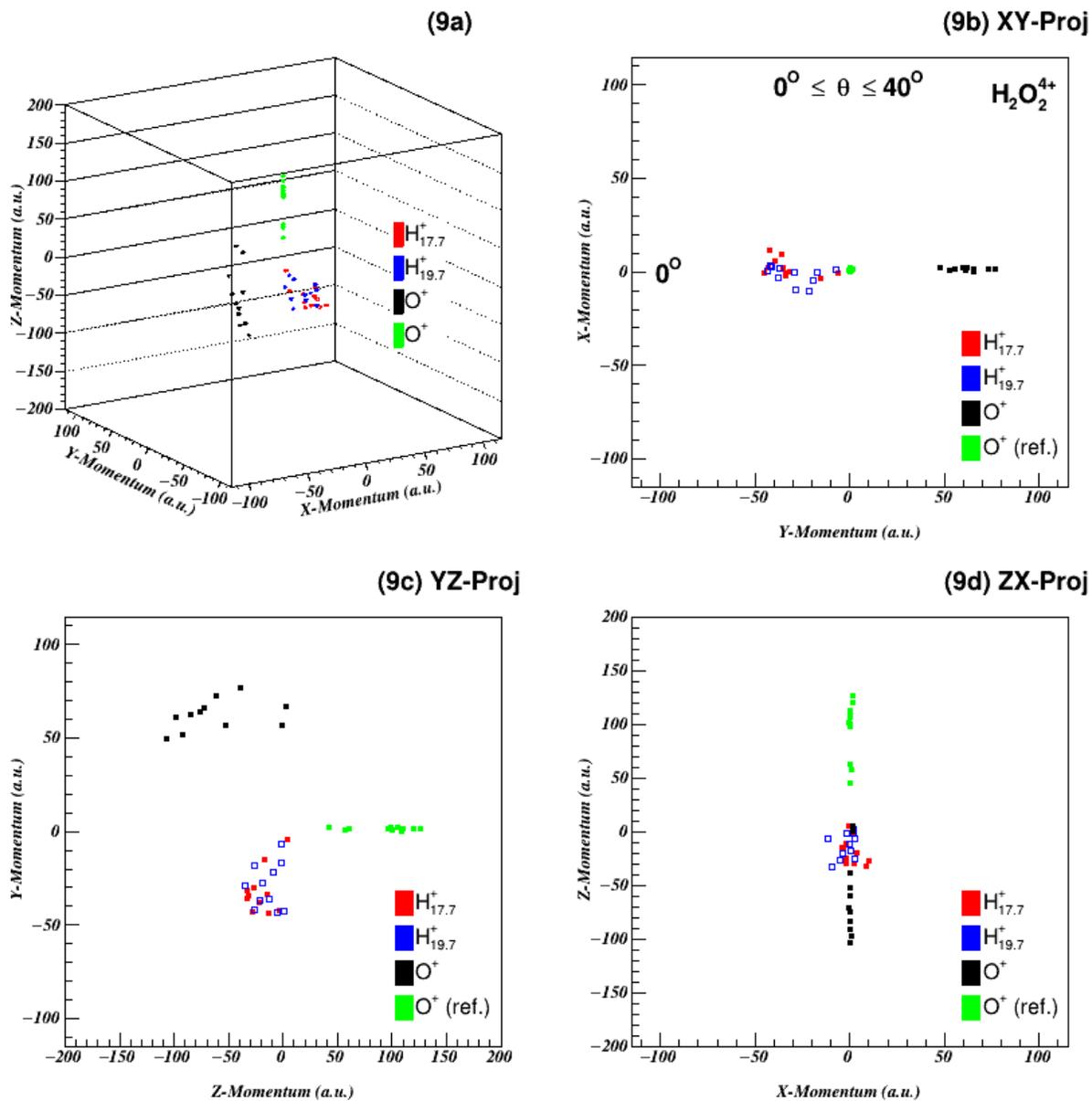

Suppl. Fig. (9): (a) Four-body Newton diagram representing the full 4-body breakup of $H_2O_2^{4+}$ for the torsional angle condition $0° \leq \theta_{H^+-H^+} \leq 40°$, which resembles the geometry close to *cis*-$H_2O_2$. The XY-, YZ-, and ZX- projections are provided in (b), (c) and (d), respectively.

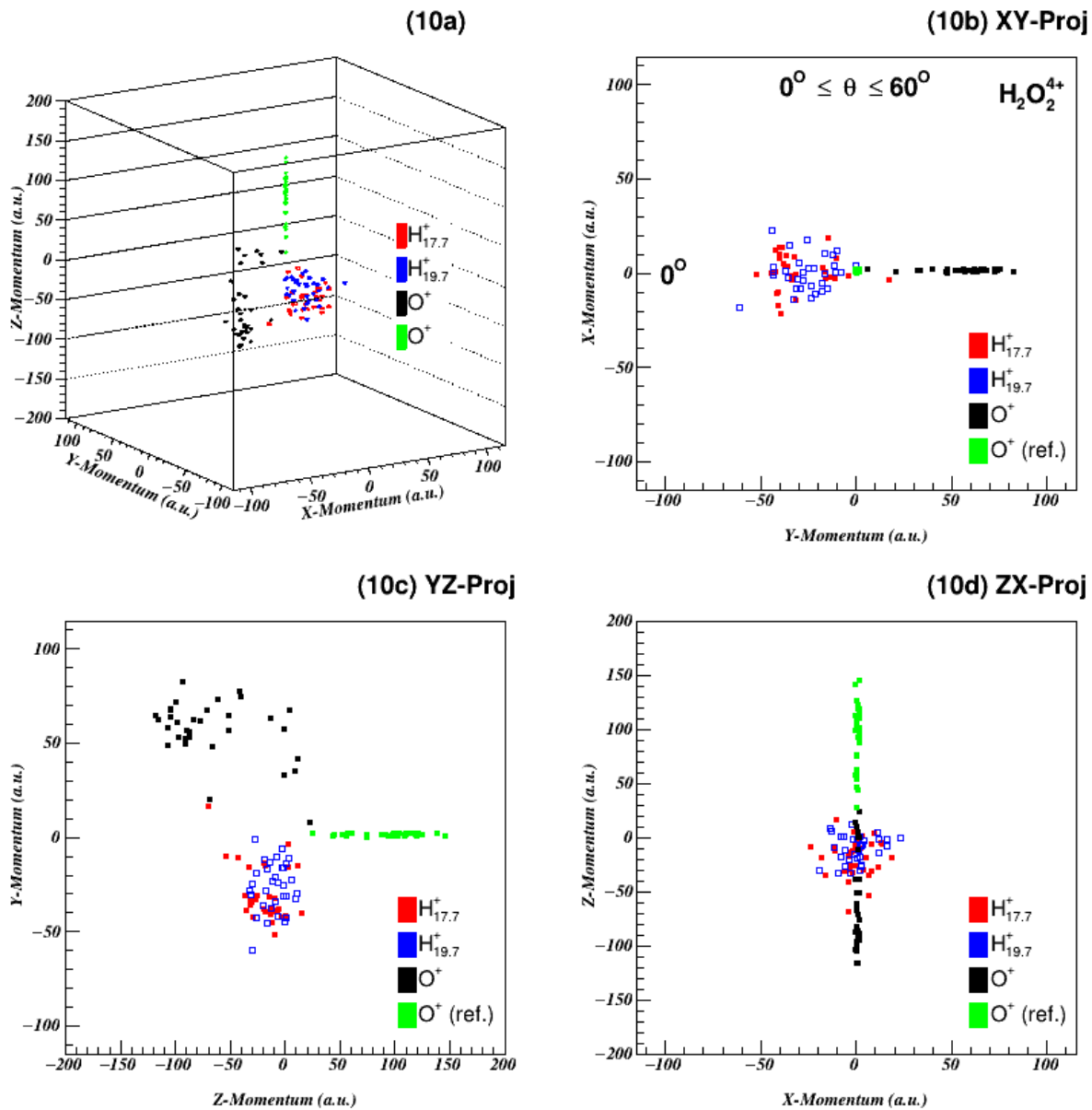

Suppl. Fig. (10): (a) Four-body Newton diagram representing the full 4-body breakup of $H_2O_2^{4+}$ for the torsional angle condition $0^o \leq \theta_{H^+-H^+} \leq 60^o$, which resembles the geometry close to *cis*-$H_2O_2$. The XY-, YZ-, and ZX- projections are provided in (b), (c) and (d), respectively.

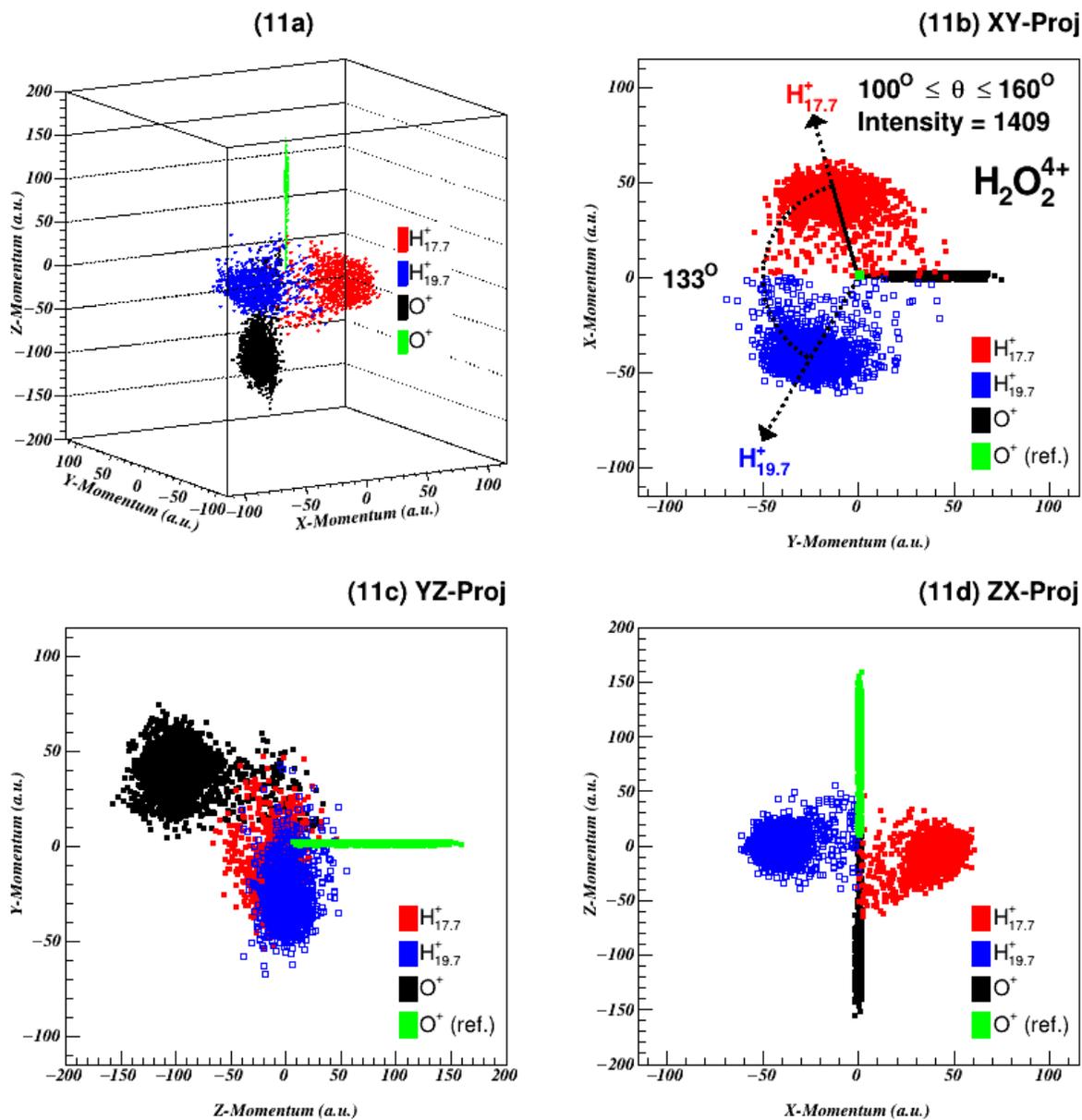

Suppl. Fig. (11): (a) Four-body Newton diagram representing the full 4-body breakup of $H_2O_2^{4+}$ for the torsional angle condition $100^o \leq \theta_{H^+-H^+} \leq 160^o$. The XY-, YZ-, and ZX-projections are provided in (b), (c) and (d), respectively. The dashed-arrow marks indicate the torsional angle 133° between the protons. As mentioned, this plot present details about one of the enantiomers of *gauche*-$H_2O_2$ depicted in the supplementary figure. (6), which corresponds to an integrated plot of two-chiral isomers (like P- and M-enantiomers).

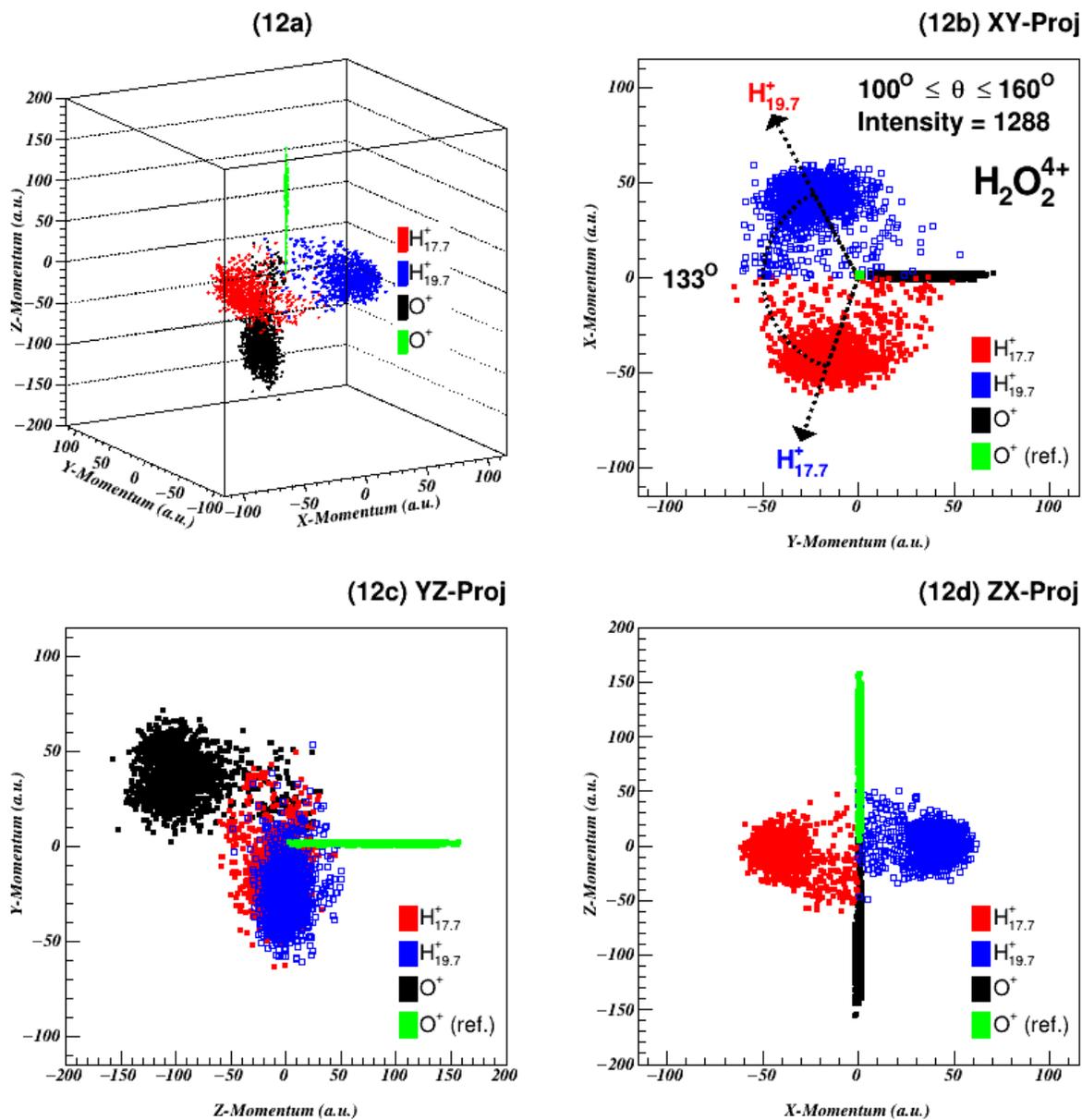

Suppl. Fig. (12): (a) Four-body Newton diagram representing the full 4-body breakup of $H_2O_2^{4+}$ for the torsional angle condition $100^o \leq \theta_{H^+-H^+} \leq 160^o$. The XY-, YZ-, and ZX-projections are provided in (b), (c) and (d), respectively. The dashed-arrow marks indicate the torsional angle 133° between the protons. As mentioned, this plot present details about one of the enantiomers of *gauche*-$H_2O_2$ depicted in the supplementary figure. (6), which corresponds to an integrated plot of two-chiral isomers (like P- and M-enantiomers).

# Acknowledgements

We thank Prof. E. Krishnakumar (RRI, Bangalore), Prof. Lokesh C. Tribedi and Prof. Vaibhav S. Prabhudesai for fruitful scientific discussions. We wish to express our gratitude to Mr. Chandan D. Bagdia who helped us while performing experiments. We also thank Mr. Apoorva D. Bhatt (TIFR, Mumbai) who has helped during the learning-phase of ROOT software.

# Author information

### Affiliations

Department of Nuclear and Atomic Physics, Tata Institute of Fundamental Research (TIFR), 1, Dr. Homi Bhabha Road, Navy Nagar, Colaba, Mumbai-400005, Maharashtra, INDIA.

**M. Nrisimha Murty, Anuvab Mandal and Deepankar Misra**

### Contributions

D.M. conceived the idea, performed experiments and contributed to data analysis apart from the interpretation of the results. M.N.M. performed the data analysis apart from contributing to the interpretation of results. A.M. performed the quantum-chemical calculations. All authors contributed to the scientific discussions and manuscript preparation.

### Competing interests

The authors declare no competing interests.

### Corresponding author

Correspondence to [Deepankar Misra](.).